\documentclass[a4paper,11pt]{article}
\pdfoutput=1
\usepackage{jcappub}
\usepackage[T1]{fontenc}
\usepackage{lmodern}
\usepackage{amsmath,amssymb,amsthm}
\usepackage{bm}
\usepackage{graphicx}
\usepackage{xcolor}
\usepackage{lettrine,Romantik}

\definecolor{revisionblue}{RGB}{0,70,180}

\usepackage{array}

\usepackage{orcidlink}
\newcolumntype{P}[1]{>{\centering\arraybackslash}p{#1}}
\newcolumntype{M}[1]{>{\centering\arraybackslash}m{#1}}
\newcommand{\be}{\begin{equation}}
\newcommand{\ee}{\end{equation}}
\newcommand{\bea}{\begin{eqnarray}}
\newcommand{\eea}{\end{eqnarray}}

\begin{document}
\title{\resizebox{\textwidth}{!}{Gauge Backreaction in Standard Model Warm Inflation}}
\author[a]{Rafid H. Dejrah\orcidlink{0000-0002-8110-296X}}
\affiliation[a]{ICTP, International Centre for Theoretical Physics, \\ Strada Costiera 11, 34151, Trieste, Italy}
\author[b]{and Anish Ghoshal\orcidlink{0000-0001-7045-302X}}
\affiliation[b]{Department of Physics and Astronomy, University of Sussex, \\ Brighton, BN1 9RH, United Kingdom}
\emailAdd{rdejrah@ictp.it}
\emailAdd{a.ghoshal@sussex.ac.uk}
\abstract{Standard-Model warm inflation (SMWI) makes QCD the microscopic origin of inflationary dissipation, so the real-time response of the gauge plasma enters the cosmological prediction. We show that a slowly relaxing gauge-helicity response cannot, in general, be absorbed into an effective friction coefficient and must instead remain dynamical. We therefore evolve the quark axial densities and a matched hard-helicity mode, with QCD sphalerons and an effective $2\leftrightarrow3$ channel generating both charge transfer and correlated stochastic sources in the full perturbation covariance system. The resulting backreaction is substantial but mainly indirect: gauge-helicity screening can reduce the warm dissipation ratio by an order-one fraction on the strong branch, while the $2\leftrightarrow3$ channel contributes only $0.2\%-0.7\%$ of the zero-density sphaleron friction directly, shifting the amplitude-normalized trajectory in both $n_s$ and $r_{\rm vac}$. The relevant trajectories, however, lack a parametrically broad hard-soft hierarchy and extend beyond strict linear response. The predictions should therefore be understood within the stated local Gaussian transport closure; a parameter-free Standard-Model (SM) result requires finite-affinity real-time $SU(3)$ drift, susceptibility, and noise kernels.}
\maketitle
\section{Introduction}
\label{sec:intro}
In the beginning, inflation need not have been cold~\cite{Starobinsky:1980te,Guth:1980zm,Linde:1981mu,Albrecht:1982wi, Sato:2015dga, Kallosh:2025ijd}. Warm inflation (WI) offers a qualitatively different realization of inflationary dynamics: the same microscopic interactions that damp the inflaton can sustain a thermal bath and source primordial fluctuations during inflation~\cite{Berera:1995ie,Berera:1996nv,Berera:1996fm,Berera:2008ar,Kamali:2023lzq}. That economy, however, comes with a stringent consistency requirement. Once dissipation is generated by an identifiable microscopic sector, the observable scalar spectrum is controlled not only by a background friction coefficient, but by the real-time response, equilibration, and stochastic fluctuations of that sector~\cite{Hall:2003zp,Graham_2009,Bastero-Gil:2011rva,Bastero-Gil:2014jsa,Laine:2021ego}. Which plasma degrees of freedom remain dynamical on cosmological time scales is therefore part of the prediction itself, not an optional refinement.

Pseudoscalar couplings to gauge fields are particularly attractive in this respect. The perturbative shift symmetry protects the inflaton potential against dangerous thermal corrections, while topological gauge fluctuations can dissipate the homogeneous motion~\cite{Berghaus:2019whh,Drewes:2023khq,Mirbabayi:2022cbt,Ito:2025lcg}. A minimal and especially predictive realization couples a quartic inflaton directly to QCD
\begin{equation}\label{eq:Lagrangian_intro}
    \mathcal{L}\supset \frac{1}{2}\left(\partial\phi\right)^2-V(\phi)-\frac{\alpha_s}{8\pi}\frac{\phi}{f}G_{\mu\nu}^a\tilde{G}^{a\mu\nu}\,,\qquad V(\phi)=\lambda\phi^4\,.
\end{equation}
In SMWI, QCD sphalerons transfer energy from the rolling inflaton to the plasma while the anomaly simultaneously transfers Chern-Simons number into quark chirality~\cite{Berghaus:2025dqi}. The induced axial chemical potentials oppose the applied topological bias; Hubble dilution and chirality-violating reactions prevent complete screening and allow a warm attractor. Recent numerical studies further show that, in phenomenologically relevant regions, the coupled WI perturbations are most reliably obtained from direct numerical evolution rather than compressed into a universal analytic expression for the scalar spectrum~\cite{Rodrigues:2025neh,ORamos:2025uqs,Kumar:2024hju}. For the SM construction itself, Ref.~\cite{ORamos:2025uqs} carried out a dedicated precision reanalysis and found quantitatively important differences in the weak- and strong-dissipation regimes.
This construction belongs to a rapidly developing sequence of QCD-based warm-inflation models. The role of light fermions in suppressing sphaleron heating, and the mechanisms by which that suppression can be alleviated, were analyzed in Ref.~\cite{Drewes:2023khq}. A complementary realization makes the SM quarks heavy during inflation so that a heavy QCD axion can sustain sphaleron heating~\cite{Berghaus:2024qca}, whereas the minimal SM construction of Ref.~\cite{Berghaus:2025dqi} instead keeps the light quarks and uses Hubble dilution to prevent complete chiral screening. Conversely, for a vanilla QCD axion with its standard shift-symmetry structure, Ref.~\cite{Zell:2024cyz} showed that the strongly dissipative warm regime is obstructed, sharpening the distinction between QCD axions and more general axion-like inflatons. Our question is orthogonal to these model-building alternatives: once a QCD-driven warm branch is present, we ask whether the gauge sector itself supplies a response mode that is too slow to be hidden inside a local friction coefficient.

A further issue arises once the gauge sector itself is allowed to respond. Ref.~\cite{Broadberry:2025ggb} emphasized that pseudoscalar interactions can induce conjugate responses for nonconserved bath quantities, and that these responses need not modify dissipation and fluctuations in the same way. For a non-Abelian plasma, the rolling $\phi\widetilde GG$ interaction can therefore bias a gauge-helicity population in addition to the fermionic axial charges. The crucial question is not whether this response can be assigned an effective chemical potential, but whether it relaxes rapidly enough to be integrated out. If a nonhydrodynamic response mode is not parametrically faster than the frequencies retained by the infrared theory, replacing it by a modified local friction coefficient discards a dynamical degree of freedom. This is the same effective-theory logic that motivates explicit kinetic variables for other long-lived populations in WI~\cite{Mirbabayi:2024eml}.

This observation is consequential for SMWI. Anomalous Langevin theory contains parity-odd non-Abelian dynamics at the magnetic scale~\cite{Akamatsu:2013pjd,Akamatsu:2014yza}; real-time studies exhibit topological drift and chirality transfer in biased plasmas~\cite{Akamatsu:2015kau,Schlichting:2022fjc}; and high-temperature QCD sphaleron and soft-axion rates continue to be refined by effective-theory and lattice calculations~\cite{Guin:2026kbp,Bouzoud:2026rur}. In addition, perturbative Landau-damping and plasmon-decay processes provide dissipation that is conceptually distinct from a helicity-affinity response~\cite{Chung:2026vnt}. These ingredients cannot in general be represented by a single replacement $\Upsilon\to\Upsilon_{\rm eff}$ without first establishing a hierarchy that removes the associated response modes.

This motivates treating SMWI as an enlarged transport problem. We retain five quark axial densities and a matched hard gauge-helicity density as explicit dynamical variables, couple them through QCD sphaleron and effective hard-helicity reaction currents, and construct the stochastic diffusion matrix from the same reaction channels that generate the deterministic charge transfer. The correlated noise is then propagated together with the inflaton and radiation perturbations through the full covariance system. In this way, deterministic backreaction and stochastic forcing are handled within the same transport framework, with the assumptions connecting them kept explicit.

The numerical results make this distinction concrete. The gauge-helicity response can screen the topological drive strongly enough to reduce the warm dissipation ratio on the strong and intermediate branches, even though the direct dissipative force from the effective helicity-changing channel remains subdominant. Consequently, the gauge response produces a coherent displacement of the amplitude-normalized inflationary trajectory, shifting $n_s$ and changing the vacuum-tensor measure $r_{\rm vac}$; at representative strong points the reduction of $Q_*$ is order unity. The effect is not simply another small friction term. Rather, the gauge response reorganizes the anomalous transport that controls the warm attractor.

The same calculation also makes clear where theoretical control begins to weaken. The relevant trajectories do not possess a parametrically wide hard-soft factorization window, several response directions are not microscopically fast compared with the expansion, and the sphaleron and perturbative affinities reach values far outside the strict linear-response regime. We consequently do not interpret the resulting cosmic microwave background (CMB) curves as parameter-free SM predictions. Instead, the robust conclusion is structural: gauge helicity is a genuine dynamical transport degree of freedom capable of materially reshaping SMWI, while a precision prediction requires finite-affinity real-time $SU(3)$ drift, susceptibility, and noise kernels. The numerical analysis serves simultaneously as a calculation of the effect within a specified local Gaussian transport closure and as a quantitative diagnosis of the microscopic information still missing.

\emph{The paper is organized as follows.}    Sec.~\ref{sec:SMWI_quark} establishes the quark-only SMWI baseline and fixes our transport conventions. Sec.~\ref{sec:enters_gauge} introduces the gauge-helicity response and its hard-soft matching, while Sec.~\ref{sec:reaction} develops the enlarged stochastic perturbation system in which the response variables remain dynamical. Sec.~\ref{sec:numerical} presents the numerical tests and the resulting primordial observables. We discuss the effective-theory interpretation and the limits of the transport description in Sec.~\ref{sec:discussion}. The appendices collect supporting derivations, perturbation-system details, and the WI slow-roll relations used in the analysis.
\section{SM-WI: quark-only baseline}
\label{sec:SMWI_quark}
For a homogeneous inflaton and a near equilibrium radiation bath
\begin{align}
    \ddot\phi+3H\dot\phi+V_{,\phi}&=-\mathcal{F}_\phi\,,\label{eq:ddot_phi_main_SMWI}\\
    \dot\rho_R+4H\rho_R&=\mathcal{P}_{\phi\rightarrow R}\,,\label{eq:dot_rho_very_main_def}\\
    3M_{\rm Pl}^2 H^2&=\frac{1}{2}\dot\phi^2+V(\phi)+\rho_R\,,\\
    \rho_R&=\frac{\pi^2}{30}g_*T^4\,.
\end{align}
Here $H\equiv \dot a/a$ is the Hubble rate, $a(t)$ is the scale factor, $\rho_R$ is the radiation energy density, $T$ is the bath temperature, $g_*$ counts the effectively relativistic energy degrees of freedom, and $M_{\rm Pl}$ is the reduced Planck mass. The quantity $\mathcal F_\phi$ denotes the dissipative force acting on the homogeneous inflaton and $\mathcal P_{\phi\rightarrow R}$ the corresponding power deposited in radiation.

When the energy transfer is local and can be represented by a friction coefficient, \(\mathcal{F}_\phi=\Upsilon\dot\phi\) and \(\mathcal{P}_{\phi\rightarrow R}=\Upsilon\dot\phi^2\). 
This equality is the leading linear-response heat source when the free energy stored in the response variables is neglected. The domain and limitation of that truncation are quantified in Eqs.~\eqref{eq:response_free_energy} and~\eqref{eq:helicity_energy_fraction}. We use
\begin{equation}\label{eq:Q_very_main_def}
    Q\equiv \frac{\Upsilon}{3H}\,.
\end{equation}
The quasi-steady radiation relation is
\begin{equation}\label{eq:quasi_steady_approx_in_SMWI}
    4H\rho_R\simeq \Upsilon\dot\phi^2\,,\qquad \rho_R\simeq \frac{3}{4}Q\dot\phi^2\,.
\end{equation}
For completeness, the WI slow-roll hierarchy underlying these background relations, together with its distinction from the microscopic transport conditions relevant to the present gauge-quark system is summarized in App.~\ref{sec:SR-WI}.

Throughout the numerical study we use \(g_*=106.75\) corresponding to the relativistic SM degrees of freedom above the electroweak scale, and the reduced Planck mass \(M_{\rm Pl}=2.435\times10^{18}\, \text{GeV}\). We write the QCD couplings as \(\alpha_s(\mu_{\rm ren})/(4\pi)\). In the one-loop six-flavor convention adopted in the numerical analysis
\begin{equation}
    \alpha_{s}^{-1}(\mu_{\rm ren})=\alpha_s^{-1}(M_Z)+\frac{11N_c-2N_f}{6\pi}\,\ln\frac{\mu_{\rm ren}}{M_Z}\,,\qquad N_c=3\,,\quad N_f=6\,.
\end{equation}
In this convention $\alpha_s\equiv g^2/(4\pi)$ is the QCD fine-structure coupling, $g$ is the gauge coupling, $M_Z$ is the reference electroweak scale used to specify the running coupling, and $N_c$ and $N_f$ denote the numbers of colors and active quark flavors. Consistently with the high-temperature SM treatment adopted here, we use the six-flavor running throughout the numerical calculation; the residual dependence on the thermal renormalization scale is assessed explicitly below. The logarithmic running matters because the sphaleron rate below scales with a high power of $\alpha_s$.

Here \(\mu_{\rm ren}\) is reserved for the renormalization scale and is not used for a chemical potential. The fiducial choice \(\mu_{\rm ren}=T\) is retained for direct comparison with the SMWI baseline, while the conventional thermal-QCD choices \(\mu_{\rm ren}=\pi T,\, 2\pi T,\, 4\pi T\) are treated in Sec.~\ref{sec:numerical} as a separate renormalization-scale sensitivity test of the observables. The coefficient of the dimension-five operator defines the associated EFT scale
\begin{equation}
    \Lambda_{\rm EFT}=\frac{8\pi f}{\alpha_s(\mu_{\rm ren})}\,.
\end{equation}
This scale tests whether the thermal bath lies below the scale inferred from the dimension-five coefficient; by itself it does not establish the validity of the full derivative expansion in a rolling background. We therefore report it together with the other validity criteria in Sec.~\ref{sec:numerical}.

At high temperature we parametrize Chern-Simons diffusion by
\begin{equation}
    \Gamma_{\rm sph}(T)=\kappa(T)\left[\alpha_s(\mu_{\rm ren})N_c\right]^5T^4\,,\qquad \Upsilon_{\rm sph}(T)=\frac{\Gamma_{\rm sph}(T)}{2Tf^2}\,.
\end{equation}
The quantity $\Gamma_{\rm sph}$ is a diffusion rate per physical volume and time and therefore has mass dimension four; $\Upsilon_{\rm sph}$ has mass dimension one, as required for a friction coefficient. The dimensionless function $\kappa(T)$ collects the nontrivial thermal-QCD normalization of the diffusion rate.
For the next-to-leading logarithmic (NLL) color-conductivity estimate we use
\begin{equation}
    \kappa(T)=1.26\frac{N_c^2-1}{(2N_c+N_f)N_c}\, W_0\left[\frac{e^{3.041}}{N_c}\sqrt{\frac{2\pi (2N_c+N_f)}{3\alpha_s(\mu_{\rm ren})}}\right]\,,
\end{equation}
where \(W_0[\,\cdots\,]\) is the principal Lambert function. This NLL expression is retained as a reproducible reference normalization; the independent high-temperature \(SU(3)\) calibration of Ref.~\cite{Guin:2026kbp} is treated as a numerical sensitivity test in Sec.~\ref{sec:numerical}.

For each slowly relaxing quark flavor, \(\mu_{5f}\) denotes the axial (chiral) chemical potential, and we use
\begin{equation}\label{eq:Here_we_chi_q}
    n_{5f}=\chi_q\mu_{5f}\,,\qquad \chi_q=\frac{N_cT^2}{3}\,.
\end{equation}
Our convention is \(\mu_{Rf}=+\mu_{5f}\) and \(\mu_{Lf}=-\mu_{5f}\). For a massless Weyl fermion, the linearized particle-minus-antiparticle density is \(N_c\mu T^2/6\); hence \(n_{5f}\equiv n_{Rf}-n_{Lf}=N_cT^2\mu_{5f}/3\). This fixes the susceptibility convention used in the affinity and charge-transfer equations. The chirality-flip rate is
\begin{equation}\label{eq:Gamma_ch_f}
    \Gamma_{{\rm ch},f}=c_{\rm ch}y_f^2 T\,,\qquad c_{\rm ch}=10^{-2}\,,
\end{equation}
for the five flavors retained after the top response is integrated out.\footnote{The top quark is included among the six active thermal flavors entering the running of $\alpha_s$, but its axial density is not retained as an independent slow response variable. Within the phenomenological chirality-flip model of Eq.~\eqref{eq:Gamma_ch_f}, its much larger Yukawa coupling makes this direction parametrically faster than the five retained flavors. A dedicated finite-temperature determination of the top axial relaxation rate is beyond the present calculation.}
Here $y_f$ is the Yukawa coupling of flavor $f$ and $c_{\rm ch}$ is the phenomenological coefficient controlling chirality relaxation. The flavor dependence of $\Gamma_{{\rm ch},f}$ is important because the slowest axial directions provide the most efficient screening of the topological drive.  The quark-only quasi-steady screening factor is
\begin{equation}\label{eq:main_def_D_q}
    D_q=\sum_f\frac{6\Gamma_{\rm sph}}{N_cT^3\left(3H+\Gamma_{{\rm ch},f}\right)}\,.
\end{equation}
This is the flavor-resolved version of the mechanism of Refs.~\cite{Drewes:2023khq, Berghaus:2025dqi}: axial charge opposes the inflaton-induced sphaleron drift while expansion and chirality violation prevent complete equilibration.

If no hard gauge-helicity variable is retained, it is convenient to write the screened topological drive as 
\begin{equation}\label{eq:equation_of_S_q_Upsilon_q}
    S_q=\frac{\dot\phi}{1+D_q}\,,\qquad \Upsilon_q=\frac{\Upsilon_{\rm sph}}{1+D_q}\,.
\end{equation}
The symbol \(S_q\) is the quark-only limit of the full quantity \(S\) defined in Eq.~\eqref{eq:main_def_S_enters_gauge}: when the hard-helicity variable \(n_h\) and the current \(J_{23}\) are switched off, \(S\rightarrow S_q\). These quantities are introduced in Sec.~\ref{sec:enters_gauge}. Thus \(S_q\) and \(S\) are not different physical forces; they denote the same topological driving term in two versions of the transport system.

\paragraph{Normalization and anomaly.} To make the factors entering the screening and stochastic sectors explicit, we use
\begin{equation}
    q(x)\equiv \frac{\alpha_s}{8\pi} G_{\mu\nu}^a\tilde{G}^{a\mu\nu}\,,
\end{equation}
and
\begin{equation}
    N_{\rm CS}(t_2)-N_{\rm CS}(t_1)=\int_{t_1}^{t_2}d^4x \,q(x)\,.
\end{equation}
For one Dirac flavor our axial-current convention is \(\partial_\mu j_{5f}^\mu=2q(x)+\,\text{chirality-violating terms}\). Consequently a unit positive topological transition changes the axial charge by \(\Delta N_{5f}=2\Delta N_{\rm CS}\). The signed current \(J_{\rm sph}\) used below is oriented oppositely to this positive topological transition, which gives the \(-2J_{\rm sph}\) source in Eq.~\eqref{eq:n_dot_5f_def}. We define the sphaleron diffusion rate by
\begin{equation}
    \Gamma_{\rm sph}\equiv\lim_{t\to \infty}\frac{\left\langle\left(\Delta N_{\rm CS}\right)^2\right\rangle}{Vt}\,.
\end{equation}
These conventions fix both the factor \(2\) multiplying each \(\mu_{5f}/T\) in Eq.~\eqref{eq:J_sph_first_appear} and the \(-2\) stoichiometric entry in the quark-charge equations. Switching off the sphaleron coupling to the hard-helicity variable, \(\zeta_h\rightarrow0\), and the hard-helicity channel, \(c_{23}\rightarrow0\), then reproduces Eq.~\eqref{eq:main_def_D_q} and the quark-only numerical scan.
\section{Gauge-helicity response}
\label{sec:enters_gauge}
We distinguish two processes that are often conflated in a scalar-friction description. The first is QCD topological diffusion. We define a signed reaction current \(J_{\rm sph}\) with linear-response coefficient \(L_s=\Gamma_{\rm sph}/2\)
\begin{equation}\label{eq:J_sph_first_appear}
    J_{\rm sph}=\frac{\Gamma_{\rm sph}}{2}\left[\frac{\dot\phi}{fT}+2\sum_f\frac{\mu_{5f}}{T}-\frac{\zeta_h}{\alpha_s}\frac{\mu_h}{T}\right]\,.
\end{equation}
The coefficient \(\zeta_h\) parameterizes the normalization with which one topological transition changes the matched hard gauge-helicity response variable. The \(1/\alpha_s\) scaling reflects the fact that a nonperturbative transition involves a collective gauge configuration containing parametrically many soft gauge quanta~\cite{Broadberry:2025ggb}. Operationally, \(2\zeta_h/\alpha_s\) is the late-time projection of one signed soft topological event onto the hard-helicity variable after matching the soft topological sector onto the hard kinetic description, and is not fixed by the anomaly alone.

The signs and coefficients in Eq.~\eqref{eq:J_sph_first_appear} can be summarized by the dimensionless thermodynamic affinity and the unscreened topological bias
\begin{equation}\label{eq:A_sph_event}
 A_{\rm sph}=\frac{\Delta W_{\rm sph}}{T}=\frac{S}{fT}\,,\qquad \xi_{\rm CS}\equiv \frac{|\dot\phi|}{fT}\,.
\end{equation}
Here \(\Delta W_{\rm sph}\) is the generalized work supplied to a positive event in the orientation defining \(J_{\rm sph}\). The equation follows from the inflaton source, the two axial-charge units transferred per flavor in our anomaly convention, and the matched hard-helicity projection. In the linear regime, \(J_{\rm sph}=L_sA_{\rm sph}\), with \(L_s=\Gamma_{\rm sph}/2>0\), and the entropy-production contribution is \(J_{\rm sph}A_{\rm sph}\geq0\). This construction fixes the relative signs without assigning an equilibrium chemical potential to Chern-Simons number. In the quark-only limit, \(S\rightarrow S_q\) of Eq.~\eqref{eq:equation_of_S_q_Upsilon_q}.

It is useful to define the affinity in velocity units
\begin{equation}\label{eq:main_def_S_enters_gauge}
    S\equiv \dot\phi+2f\sum_f\mu_{5f}-\zeta_h\frac{f}{\alpha_s}\mu_h\,,
\end{equation}
so that 
\begin{equation}\label{eq:J_sph_relation}
    J_{\rm sph}=f\Upsilon_{\rm sph}S\,.
\end{equation}
The corresponding force on the homogeneous inflaton is \(\mathcal{F}_{\rm sph}=J_{\rm sph}/f=\Upsilon_{\rm sph}S\).
The current $J_{\rm sph}$ has mass dimension four and counts signed topological reaction activity; the combination $S$ has the dimensions of an inflaton velocity. Thus $S/\dot\phi$ directly measures the fraction of the bare rolling-field drive that survives quark and gauge screening.

The second process is the perturbative non-Abelian helicity-changing channel identified schematically in Ref.~\cite{Broadberry:2025ggb}. We write
\begin{equation}\label{eq:main_J_23_def}
    J_{23}=L_{23}\left[\alpha_s\frac{\dot\phi}{fT}-\frac{\mu_h}{T}\right]\,,\qquad L_{23}=c_{23}\alpha_s^3T^4\,.
\end{equation}
The corresponding dimensionless perturbative affinity is
\begin{equation}\label{eq:A_23_main_def}
    A_{23}\equiv \alpha_s\frac{\dot\phi}{fT}-\frac{\mu_h}{T}\,.
\end{equation}
The coefficient $L_{23}$ is the Onsager-like mobility of the effective hard-helicity channel and $c_{23}$ is its dimensionless matching coefficient. The notation $2\leftrightarrow3$ labels the effective number-changing hard process; it should not be read as a claim that one uniquely identified vacuum scattering amplitude exhausts the thermal collision kernel.

This channel is important conceptually even when its direct scalar noise is small: it is the process that relaxes gauge helicity toward the applied perturbative affinity. The force conjugate to the rolling pseudoscalar is
\begin{equation}
    \mathcal{F}_{23}=\frac{\alpha_s}{f}J_{23}\,.
\end{equation}
We emphasize that \(J_{23}\) is a phenomenological perturbative hard-helicity relaxation channel, not a calculated finite-temperature \(SU(3)\) collision integral. It is intended to collect number-changing non-Abelian processes built from the cubic and quartic gauge vertices, schematically  \(gg\leftrightarrow ggg\) and crossed channels, for which dimensional collision-rate counting gives an event activity of order \(\alpha_s^3 T^4\). The coefficient \(c_{23}\) absorbs the unresolved screening, Bose-enhancement, collinear and possible Landau-Pomeranchuk-Migdal effects, as well as the precise helicity projection. Accordingly, varying \(c_{23}\) tests sensitivity to an unknown matching coefficient; it is not a probability distribution or a first-principles error estimate. Ref.~\cite{Broadberry:2025ggb} motivates the affinity structure but does not determine the numerical \(SU(3)\) collision coefficient. The label \(2\leftrightarrow3\) is therefore mnemonic; elastic redistribution, collinear splitting, soft exchange, and overlap with other perturbative channels can renormalize the same effective hard-helicity mobility.

The same non-Abelian expansion contains \(A^3\dot\phi/f\) and \(A^4\dot\phi/f\) terms that are not exhausted by a chemical-potential interpretation~\cite{Broadberry:2025ggb}. We reserve
\begin{equation}
    \Upsilon_{\rm dir}=C_{\rm dir}\frac{\alpha_s^5T^3}{f^2}\,,
\end{equation}
for this additional local contribution, whose matching is not fixed here. The fiducial calculation sets \(C_{\rm dir}=0\). This is a conservative separation of known thermodynamic structure from an unknown \(SU(3)\) matching coefficient, not an assertion that the direct channel vanishes.

The coefficient \(C_{\rm dir}\) should not be interpreted as spanning every omitted gauge contribution. In particular, the perturbative Landau-damping and plasmon-decay terms isolated in Ref.~\cite{Chung:2026vnt} are conceptually distinct from the chemical-affinity response above, while genuinely nonlocal memory effects require a frequency-dependent kernel rather than a local \(\Upsilon_{\rm dir}\). Setting \(C_{\rm dir}=0\) thus defines the fiducial calculation.

\subsection{Operational definition and matching of hard gauge helicity}
The gauge response variable is not an ordinary chemical potential for an exactly conserved global charge. We define $n_h$ as a matched kinetic variable measuring the occupation-number difference between the two transverse helicities of hard gluonic quasiparticles. Introducing a hard-soft separation scale $\Lambda_{\rm sep}$ with $g^2T\ll gT\ll\Lambda_{\rm sep}\lesssim T$
\begin{equation}
n_h\equiv d_A\!\int_{p\gtrsim\Lambda_{\rm sep}}\!\frac{d^3p}{(2\pi)^3}[f_+(p)-f_-(p)]\,,\qquad d_A\equiv N_c^2-1\,.
\end{equation}
Here $f_\pm(p)$ are matched distribution functions for positive- and negative-helicity transverse hard gluonic quasiparticles, $p$ is the momentum magnitude, $N_c=3$, $d_A=8$, $g$ is the QCD gauge coupling, and $T$ is the temperature. The construction presumes a fixed gauge and quasiparticle/Wigner-function prescription for the hard modes. The lower limit \(p\gtrsim \Lambda_{\rm sep}\) is deliberate: \(n_h\) is the retained hard-quasiparticle helicity moment, while magnetic-scale configurations belong to the Chern-Simons sector. Extending the same integral over all momenta without simultaneously rematching the soft sector would double count the degrees of freedom that generate topological diffusion. This does not assert that only hard modes can carry helicity;  it defines the hard variable used in the matched description. Neither $n_h$ nor $\mu_h$ is claimed to be an exact gauge-invariant conserved charge. Dependence on the separation scale and on the hard-mode prescription must cancel against the matched susceptibility, the hard-helicity change associated with each topological event, and the relaxation coefficient in a complete real-time calculation; in the present calculation that dependence is represented by $c_\chi$, $\zeta_h$, and $c_{23}$.

The reduction to a single \(n_h\) is itself a moment closure of a kinetic description. It assumes that momentum- and angle-dependent distortions orthogonal to the retained helicity moment relax sufficiently rapidly that they need not be resolved on cosmological time scales. If additional helicity moments are long lived, the infrared theory must be enlarged further to evolve the corresponding phase-space distributions \(f_{\pm}(p,\mathbf{x},t)\). Thus \(n_h\) is the minimal response variable retained in the present calculation; it is not an assertion that an arbitrary nonequilibrium gauge distribution is characterized by one chemical potential.

Near the unbiased reference state one may introduce the generating distributions
\begin{equation}
f_\pm(p)=\left[\exp\!\left(\frac{\omega_p\mp\mu_h}{T}\right)-1\right]^{-1}.
\end{equation}
The name ``chemical potential'' is used only in this kinetic-response sense: non-Abelian gauge helicity is not an exactly conserved particle number. The displayed Bose form is used only to define the derivative at $\mu_h=0$; it is not extrapolated as a physical finite-density distribution when $|\mu_h|/T\gtrsim1$. For a literal cutoff definition, the reference susceptibility is
\begin{equation}
\chi_h^{(0)}(\Lambda_{\rm sep})
 =\frac{2d_A}{T}\int_{p\gtrsim\Lambda_{\rm sep}}
 \frac{d^3p}{(2\pi)^3}f_B(1+f_B)
 =\frac{d_AT^2}{3}\,F_\chi\!\left(\frac{\Lambda_{\rm sep}}{T}\right),
\end{equation}
where $f_B=[\exp(p/T)-1]^{-1}$ and
\(F_\chi(x)=(3/\pi^2)\int_x^\infty dy\,y^2e^y/(e^y-1)^2\), with $F_\chi(0)=1$. A literal Bose distribution at finite \(\mu_h\) would additionally require $|\mu_h|<\min_{p\gtrsim\Lambda_{\rm sep}}\omega_p$; the large-affinity benchmarks do not establish this condition and are therefore interpreted only through the linearized response variable. To avoid introducing an uncomputed cutoff function into the present numerical treatment, we absorb $F_\chi$, interaction corrections, dispersion effects, and scheme dependence into one matched coefficient and define
\begin{equation}\label{eq:chi_h_matched}
\chi_h=c_\chi\frac{d_A}{3}T^2\,.
\end{equation}
\(F_\chi(\Lambda_{\rm sep}/T)\) is the explicit free-theory cutoff factor associated with the literal hard integral, whereas \(c_\chi\) is the matching coefficient used in the numerical calculation. They coincide only after choosing a particular matching prescription; they are not independent estimates of the same numerical quantity. Thus $c_\chi=1$ denotes the uncut free reference normalization, not a claim that the literal hard-cutoff susceptibility equals its all-momentum value. Changing the factorization prescription requires rematching $c_\chi$, $c_{23}$, and $\zeta_h$ together.

\paragraph{Thermodynamic energy stored in the response variable.} The linear susceptibility defines the quadratic free-energy cost of the retained response variables
\begin{equation}\label{eq:response_free_energy}
 \Delta\mathcal F_{\rm resp}
 =\frac{1}{2}\sum_f\chi_q\mu_{5f}^2
 +\frac{1}{2}\chi_h\mu_h^2
 +\mathcal O(\mu^4)\,.
\end{equation}
In the strict linear-response expansion, this correction is second order in the affinities, whereas the drift equations are retained only to first order. It is therefore consistently omitted from the central linearized Friedmann and radiation equations only when the response biases are small. A useful measure obtained from Eq.~\eqref{eq:chi_h_matched} is
\begin{equation}\label{eq:helicity_energy_fraction}
 \frac{\Delta\mathcal F_h}{\rho_R}
 \simeq\frac{5c_\chi d_A}{\pi^2g_*}
 \left(\frac{\mu_h}{T}\right)^2
 \simeq0.038\,c_\chi
 \left(\frac{\mu_h}{T}\right)^2
 \quad(d_A=8,\ g_*=106.75)\,.
\end{equation}
This estimate is itself controlled only for \(\lvert\mu_h\rvert/T\ll1\). Its growth to order unity on the strongest benchmarks is therefore not used as a quantitative correction to the background; rather, it is an additional indication that those points lie outside the regime in which the linearized equation of state is controlled.  A controlled inclusion of \(\Delta\mathcal F_{\rm resp}\), its pressure, and their perturbations would require the corresponding nonlinear susceptibility at finite affinity, which is presently unknown.   Consequently, the strong-affinity trajectories are retained only to diagnose the breakdown of linear response, and no uncomputed response energy is inserted into the existing numerical results.

\paragraph{Invariance  under a change of hard-helicity normalization.}
A redefinition of the hard-helicity variable, $n_h'=\tilde a\, n_h$, cannot change physics. Invariance of the work term $\mu_hn_h$ requires $\mu_h'=\mu_h/\tilde a$, and hence
\begin{equation}
\chi_h'=\tilde a^{\,2}\chi_h\,,\qquad \nu_{r,h}'=\tilde a\,\nu_{r,h}\,.
\end{equation}
The nonzero constant $\tilde a$ changes only the normalization convention, while $\nu_{r,h}$ denotes the hard-helicity change per event of reaction channel $r$. Thus $\zeta_h'=\tilde a\,\zeta_h$, and the perturbative event normalization transforms identically. Since each reaction channel $r$ contributes $D^{(r)}_{ij}=2L_r\nu_{r,i}\nu_{r,j}$ to the diffusion matrix, the deterministic equations and covariance transform as tensors under this change of normalization. The free-energy cost, physical relaxation spectrum, and curvature perturbation are invariant. Our convention fixes this redundancy through Eq.~\eqref{eq:chi_h_matched}; variation of $c_\chi$ then probes matching uncertainty rather than arbitrary normalization.

We test this covariance directly in the numerical implementation. Under the transformations \(n_h\rightarrow \tilde a\,n_h\) with \(\tilde a=1/3,\,3,\) and \(10\), together with the corresponding transformations of \(\mu_h,\, \chi_h,\) the hard-helicity change assigned to each event, and the covariance tensor, the final curvature power changes by at most \(3.3  \times 10^{-11}\). The test therefore verifies that the normalization of the matched hard-helicity variable is conventional rather than a physical source of the reported effect.

The hard-helicity current \(J_{23}\) and soft sphaleron current \(J_{\rm sph}\) are therefore distinct matched channels. The coefficient \(\zeta_h\) specifies how a topological event is mapped onto the hard-helicity variable. A first-principles computation must determine the \(\Lambda_{\rm sep}\) dependence, susceptibility, and the projection of a topological event onto the hard-helicity variable in a common real-time scheme; the present calculation keeps this matching dependence explicit.

For the literal cutoff definition we additionally impose the Bose-positivity check
\begin{equation}
    B_h(N)\equiv\frac{\left|\mu_h(N)\right|}{\omega_{\rm min}\left(\Lambda_{\rm sep}, T\right)}\,.
\end{equation}
For massless hard quasiparticles, \(\omega_{\rm min}\simeq \Lambda_{\rm sep}\). Hence \(B_h\geq 1\) excludes interpreting the generating Bose distribution as a literal Bose distribution at that finite value of \(\mu_h\), while it does not by itself invalidate \(n_h\) as a linear matched response variable.

Eq.~\eqref{eq:main_J_23_def} can be written as a relaxation law with
\begin{equation}
    \Gamma_h\equiv \frac{L_{23}}{T\chi_h}=C_h\alpha_s^3T\,,\qquad C_h=\frac{3c_{23}}{c_\chi d_A}\,.
\end{equation}
For \(N_c=3\) and the fiducial choice \(c_{23}=c_{\chi}=1\), one has \(C_h=3/8\).
\subsection{Quasi-steady reduction}
The background is solved in the quasi-steady response regime, but the perturbations retain the charge variables dynamically. Setting this in Eqs.~\eqref{eq:n_dot_5f_def} and~\eqref{eq:n_dot_h_def} gives
\begin{equation}\label{eq:2f_sum_D_q}
    2f\sum_f\mu_{5f}=-D_qS\,,
\end{equation}
where \(D_q\) is defined in Eq.~\eqref{eq:main_def_D_q} and 
\begin{equation}\label{eq:quasti_approx_3H_Js}
    3H\chi_h\mu_h=\frac{2\zeta_h}{\alpha_s}J_{\rm sph} +J_{23}\,.
\end{equation}
Solving these equations together with Eqs.~\eqref{eq:J_sph_relation} and~\eqref{eq:main_J_23_def} yields a compact gauge-screening parameter\footnote{Cf. App.~\ref{sec:der_quasi_steady_gauge} for detailed discussion about this expression.}
\begin{equation}\label{eq:main_def_D_h}
    D_h=\frac{\zeta_h^2\Gamma_{\rm sph}}{\alpha_s^2T\chi_h(3H+\Gamma_h)}\,.
\end{equation}
The net sphaleron affinity becomes
\begin{equation}\label{eq:S_dot_phi_ratio}
    \frac{S}{\dot\phi}=\frac{1-\zeta_h\Gamma_h/\left(3H+\Gamma_h\right)}{1+D_q+D_h}\,.
\end{equation}
For the central \(\zeta_h=1\) this simplifies to
\begin{equation}\label{eq:simplieifed_total_affinity_R_h_def}
    \frac{S}{\dot\phi}=\frac{1}{\left(1+R_h\right)\left(1+D_q+D_h\right)}\,,\qquad R_h\equiv \frac{\Gamma_h}{3H}\,.
\end{equation}
The induced gauge chemical potential is obtained from Eq.~\eqref{eq:quasti_approx_3H_Js}.
Equation~\eqref{eq:simplieifed_total_affinity_R_h_def} makes the screening mechanism transparent. $D_q$ measures feedback through quark axial charge, $D_h$ measures feedback through the hard gauge-helicity response, and $R_h$ compares hard-helicity relaxation with Hubble dilution. The key point is that $D_h$ and $R_h$ probe different ratios: $D_h$ can be appreciable even when $R_h\ll1$. A slow gauge response can therefore accumulate enough bias to screen the sphaleron current without approaching instantaneous microscopic equilibration.

The total chemical drag entering the background is
\begin{equation}
    \Upsilon_{\rm chem}=\Upsilon_{\rm sph}\frac{S}{\dot\phi}+\Upsilon_{23}\,,
\end{equation}
where
\begin{equation}
    \Upsilon_{23}\equiv \frac{\alpha_sJ_{23}}{f\dot\phi}\,.
\end{equation}
Thus the \(2\leftrightarrow3\) channel has two logically distinct effects: it changes \(\mu_h\) and hence screens the topological affinity, but it also produces its own dissipative force. The complete background coefficient is therefore 
\begin{equation}\label{eq:Upsilon_tot}
    \Upsilon_{\rm tot}=\Upsilon_{\rm sph}\frac{S}{\dot\phi}+\frac{\alpha_sJ_{23}}{f\dot\phi}+C_{\rm dir}\frac{\alpha_s^5T^3}{f^2}\,.
\end{equation}
This is the total net local dissipation coefficient used in the gauge+quark background. In the fiducial calculation \(C_{\rm dir}=0\); the last term is displayed to distinguish the screened sphaleron contribution, the dissipative force associated with \(J_{23}\), and additional local gauge-field dissipation not fixed by the present matching.

The dimensionless quantities \(\xi_{\rm CS},\, A_{23},\, R_h,\) and \(D_h\) separate four distinct questions: the bare topological drive, the actual perturbative helicity bias, microscopic hard-helicity relaxation relative to expansion, and feedback of the hard response on the sphaleron current. They are evaluated numerically in Sec.~\ref{sec:numerical}, where no instantaneous gauge-equilibration assumption is made.

\paragraph{Timescale test.} Define the homogeneous response vector and susceptibility matrix by
\begin{equation}
    \bm{q}=\left(n_{5u},\, n_{5d},\, n_{5s},\, n_{5c},\, n_{5b},\, n_h\right)^T\,,\qquad \bm{\chi}=\text{diag}\left(\chi_q,\, \chi_q,\, \chi_q,\, \chi_q,\,\chi_q,\,\chi_h\right)\,.
\end{equation}
Linearizing the reaction parts of Eqs.\eqref{eq:n_dot_5f_def} and~\eqref{eq:n_dot_h_def} about the instantaneous background gives schematically
\begin{equation}
    \bm{\dot{q}}=-\left(3H \bm{1}+ \bm{R}_{\rm rxn}\right)\bm{q}+\bm{s}\dot\phi\,,\qquad \bm{R}_{\rm rxn}\equiv -\frac{\partial \bm{\dot{q}}_{\rm rxn}}{\partial \bm{q}}\,,\quad \bm{s}\equiv \frac{\partial\bm{\dot{q}}_{\rm rxn}}{\partial\dot\phi}\,.
\end{equation}
If \(\gamma_i\) is a stable instantaneous eigenvalue of the relevant response generator, an algebraic elimination requires
\begin{equation}\label{eq:quasiadiabatic_approx}
    \frac{H}{|\gamma_i|}\ll 1\,,\qquad\frac{|\dot\gamma_i|}{|\gamma_i|^2}\ll 1\,.
\end{equation}
The first condition requires relaxation within a small fraction of a Hubble time; the second requires the rate itself to vary little during one relaxation time. Together, these conditions express the required quasi-adiabatic separation between the response mode and the cosmological evolution. The relaxation spectrum of the reaction sector and the full homogeneous evolution are presented in Sec.~\ref{sec:numerical}. We do not use the quasi-steady background radiation as an argument that the response sector has been integrated out of the low-energy description. The perturbations retain all axial and hard-helicity response variables dynamically. The algebraic reduction is used only for the homogeneous background after direct comparison with the corresponding full charge evolution. This distinction is essential when a response time is comparable to the cosmological time scale.
\section{Stochastic fluctuations from reaction channels}
\label{sec:reaction}
If all nonhydrodynamic bath variables relax rapidly, their effect on the inflaton can be summarized by local constitutive coefficients and an associated  fluctuation kernel. In the parameter range of interest, the required separation of time scales is not parametrically satisfied. We therefore evolve the response densities alongside the inflaton and radiation perturbations. Chemical potentials enter only as variables conjugate to those densities; they are not inserted as external parameters into the conventional inflaton-radiation perturbation equations. The result is a low-energy transport description with explicit slow response variables, rather than a modification that changes only local coefficients in the usual WI system.

The stochastic equations below are Markovian only in the effective-theory sense: we assume that the microscopic correlation time of the reaction currents is short compared with \(H^{-1}\) and with the periods of the scalar modes retained in the calculation, so that the nonlocal memory kernel may be replaced by time-local drift and delta-correlated reaction noise. The displayed equations are time local because of this approximation;  their locality is not itself evidence that the microscopic plasma is Markovian. We do not independently compute the microscopic correlation time \(\tau_{\rm corr}\). Establishing \(\omega\,\tau_{\rm corr}\ll1\), in particular for modes near horizon crossing, requires the frequency-dependent real-time QCD response and noise kernels that are not presently available. Markovianity is therefore an assumption of the local transport description rather than a hierarchy established by the numerical calculation.

We normalize the local stochastic reaction-current  fluctuations in physical time by
\begin{equation}\label{eq:event_noise_physical}
\left\langle\delta J_r(t,{x})\,\delta J_s(t',{x}')\right\rangle
 =2L_r\,\delta_{rs}\,\delta(t-t')\,\delta^{(3)}({x}-{x}').
\end{equation}
With the Fourier convention \(X(\textbf{x})=\int d^3k\, (2\pi)^{-3}\, e^{i\textbf{k}.\textbf{x}}\, X_{\textbf{k}}\) and  \(dN=Hdt\), this becomes
\begin{equation}\label{eq:event_noise_efold}
\left\langle\delta J_{r,{k}}(N)\,\delta J_{s,{k}'}(N')\right\rangle
 =(2\pi)^3\delta^{(3)}({k}+{k}')\,
 \frac{2L_rH}{a^3}\,\delta_{rs}\,\delta(N-N').
\end{equation}
The factors of \(a^{-3}\) and \(H\) arise, respectively, from the physical-to-comoving spatial delta function and the conversion of the time delta function. 

For reference, the complete homogeneous system solved before linearization is
\begin{equation}\label{eq:complete_system_before_linear}
    \begin{aligned}
        \ddot\phi+3H\dot\phi+V_{,\phi}&=-\left(\frac{J_{\rm sph}}{f}+\frac{\alpha_sJ_{23}}{f}+\Upsilon_{\rm dir}\dot\phi\right)\,,\\
        \dot\rho_R+4H\rho_R&=\dot\phi\left(\frac{J_{\rm sph}}{f}+\frac{\alpha_s J_{23}}{f}+\Upsilon_{\rm dir}\dot\phi\right)\,,\\
        \dot{n}_{5f}+3Hn_{5f}&= -2J_{\rm sph}-\Gamma_{{\rm ch},f} n_{5f}\,,\\
        \dot{n}_h+3Hn_h&= \frac{2\zeta_h}{\alpha_s}J_{\rm sph}+J_{23}\,,\\
        3M_{\rm Pl}^2H^2&=\frac{1}{2}\dot\phi^2+V+\rho_R\,.
    \end{aligned}
\end{equation}
The currents, susceptibilities, and reaction rates appearing in Eq.~\eqref{eq:complete_system_before_linear} are those defined in Secs.~\ref{sec:SMWI_quark} and~\ref{sec:enters_gauge}. This is the system from which the background and perturbation drift matrix are generated; no second phenomenological dissipation equation is solved in parallel.

The numerical perturbation basis is ordered as
\begin{equation}\label{eq:X_k_u_defs}
    {\bm X}_k=\left(\delta\phi,\, u,\, \delta\rho_R,\, {\bm \Psi},\, \delta n_{5u},\, \delta n_{5d},\, \delta n_{5s},\, \delta n_{5c},\, \delta n_{5b},\,\delta n_{h}\right)^T\,, \qquad u\equiv \frac{d\,\delta\phi}{dN}\,.
\end{equation}
The variable $u=d\delta\phi/dN$ is the $e$-fold derivative of the inflaton perturbation; $N\equiv\ln a$ is used as the time coordinate in the mode evolution. The matrix ${\bm A}_k$ below is the linear drift matrix, ${\bm B}_k$ maps independent unit white-noise variables ${\bm\xi}_k$ into the state, and ${\bm\Sigma}_k\equiv\langle{\bm X}_k{\bm X}_k^\dagger\rangle$ is the equal-time covariance matrix. This deterministic covariance evolution is conceptually related to the moment-based treatment implemented in WI2easy~\cite{Rodrigues:2025neh} and to the stabilized correlation-matrix formulation developed in DSWIM~\cite{Kumar:2026eak}.

Here \(\bm{\Psi}\) is the scalar radiation-momentum potential. We use the spatially flat scalar slicing; the lapse and expansion perturbations are eliminated algebraically before the drift matrix is formed. The corresponding constraint equations, curvature-extraction vector, and stochastic source are collected in App.~\ref{sec:perturbations}.

Linearization gives
\begin{equation}
    \frac{d{\bm X}_k}{dN}={\bm A}_k(N){\bm X}_k+{\bm B}_k(N){\bm \xi}_k\,,
\end{equation}
with unit white noise variables \({\bm \xi}_k\). Rather than sampling many stochastic realizations, we evolve the equal-time covariance
\begin{equation}
    \frac{d{\bm \Sigma}_k}{dN}={\bm A}_k{\bm\Sigma}_k+{\bm\Sigma}_k{}{\bm A}_k^\dagger+{\bm B}_k{\bm B}_k^\dagger\,.
\end{equation}
Numerical stochastic evolution of WI perturbations has also been implemented directly in the Stochastic Warm Inflation Module (SWIM) framework for generating and analyzing warm-inflation power spectra~\cite{Kumar:2026mvz}. Deterministic correlation-matrix evolution provides the complementary formulation; a numerically stabilized implementation, including correlated thermal noise through the diffusion matrix, was recently developed in DSWIM~\cite{Kumar:2026eak}. For a reaction channel \(r\), let \(\bm{g}_{r,k}\) denote the source vector by which a unit reaction current enters the physical-time evolution of the state in Eq.~\eqref{eq:X_k_u_defs}. The \(e\)-fold diffusion contribution is then
\begin{equation}\label{eq:mixed_state_diffusion}
 {\bm B}_{r,k}{\bm B}_{r,k}^{\dagger}
 =\frac{2L_r}{a^3H}\,{\bm g}_{r,k}{\bm g}_{r,k}^{\dagger}\,.
\end{equation}
This construction preserves the correlations generated by the same microscopic reaction across all perturbed variables, rather than assigning independent noise sources to the different components of the state vector.

\paragraph{Equilibrium consistency and the finite affinity.}
In the following reaction-space notation, ${\bm\nu}_{r}$ is the stoichiometric vector specifying the change of each retained density in one positive event of reaction channel $r$, ${\bm R}$ is the linear relaxation matrix of the chemical-density subblock, ${\bm C}_{\rm eq}$ is its equilibrium covariance, and ${\bm D}$ is the corresponding diffusion matrix. In a homogeneous density basis, linear irreversible thermodynamics gives
\begin{equation}
    \mathbf{R}=\sum_{r}\frac{L_{r}}{T}{\bm \nu}_{r}{\bm \nu}_{r}^T{\bm \chi}^{-1}\,,\qquad {\bm C}_{\rm eq}=T{\bm \chi}\,, \qquad \mathbf{D}=\sum_{r}2L_{r}\bm{\nu}_{r}{\bm\nu}_{r}^T\,,
\end{equation}
so that
\begin{equation}
    {{\bm R}{\bm C}_{\rm eq}+{\bm C}_{\rm eq}{\bm R}^{T}={\bm D}}\,.
\end{equation}
This identity concerns the unbiased chemical-density reaction subblock; it does not assert global equilibrium for the expanding inflaton-radiation system. For each reaction channel \(r\), let \(R^+_r\) and \(R_r^-\) be the forward and backward event rates per unit physical volume and unit physical time, and let \(A_r\equiv \Delta W_r/T\) be the dimensionless thermodynamic affinity in the orientation chosen for the positive event. At finite affinity, the net current and total reaction rate are independent microscopic quantities
\begin{equation}
    J_{r}=R_r^+-R_r^-\,,\qquad \mathcal{N}_{r}=R_r^++R_r^-\,.
\end{equation}
Near equilibrium \(J_{r}\simeq L_{r}A_{r}\) and \(\mathcal{N}_{r}\simeq2L_{r}\), but the latter relation is not fixed by the linear-response coefficient once \(|A_{r}|\gtrsim1\). Our covariance calculation therefore extrapolates the linear drift and Gaussian noise prescription together beyond the small-affinity regime. The resulting spectra test the sensitivity to that approximation; they are not a first-principles determination of the QCD noise kernel at finite affinity.

\paragraph{Energy conservation for stochastic reaction sources.} For the sphaleron channel \(2L_s=\Gamma_{\rm sph}\), so one common stochastic current produces correlated entries in the inflaton radiation, axial, and hard-helicity equations. Its radiation component is fixed by energy conservation for the same stochastic reaction current rather than by an independent white-noise amplitude. Denoting by \(\mathcal{S}_{\rho_i}^{\rm sph}\) the contribution of the stochastic sphaleron event current to the physical-time evolution equation for the corresponding energy density, the source terms obey
\begin{equation}\label{eq:all_S_s}
    \mathcal{S}_{\rho_\phi}^{\rm sph}+\mathcal{S}_{\rho_R}^{\rm sph}+\mathcal{S}_{\rho_{\rm resp}}^{\rm sph}=0\,.
\end{equation}
Eq.~\eqref{eq:all_S_s} is an identity among the stochastic source terms, not the complete perturbed cosmological continuity equation; the deterministic Hubble, metric, momentum-transfer, and constraint terms remain in the full system. At the retained linear-response order the response free energy is second order in the affinities, so \(\mathcal{S}_{\rm resp}^{\rm sph}=0\) and the radiation source is fixed by \(\mathcal{S}_{\rho_R}^{\rm sph}=-\mathcal{S}_{\rho_\phi}^{\rm sph}\). An extension beyond linear response would require the nonlinear response equation of state and the corresponding noise kernel.

In the state ordering of Eq.~\eqref{eq:X_k_u_defs}, a sphaleron event changes each retained axial density by \(-2\) units and the hard-helicity variable by \(2\zeta_h/\alpha_s\) units. The effective \(2\leftrightarrow 3\) channel has \(L_{23}=c_{23}\alpha_s^3T^4=\Gamma_h\chi_hT\) and changes \(n_h\) by one unit in the adopted kinetic normalization. Chirality-flip currents are independent between flavors in the present model. These reaction vectors determine the complete diffusion matrix.

It is useful to compare the size of the reaction currents with their direct forcing of the inflaton. Along the fiducial scan, the effective $2\leftrightarrow3$ channel has substantial reaction activity, while its corresponding inflaton-force variance carries an additional coupling suppression and remains at the sub-percent level relative to the sphaleron force variance. The perturbative channel can therefore be important for setting \(\mu_h\) without dominating the direct scalar noise.

\paragraph{Initial covariance.} Every mode in the numerical scan is initialized on a mode-dependent subhorizon surface satisfying
\begin{equation}
    \left.\frac{k}{aH}\right|_{\rm in}=100\,.
\end{equation}
The inflaton block is generated from a single Bunch-Davies mode function In the dimensionless numerical basis \((\delta\phi,\, u)\) it is
\begin{equation}
    {\bm \Sigma}_{\rm BD}=\frac{1}{2ka^2}\begin{pmatrix}
        1 & -1+ix_{\rm in}\\-1-ix_{\rm in}& 1+x_{\rm in}^2
    \end{pmatrix}\,, \qquad x_{\rm in}\equiv \frac{k}{aH}\,.
\end{equation}
The fiducial calculation sets the initial radiation, axial, and hard-helicity covariance blocks, and all initial cross-correlations with them, to zero. Thermal and chemical fluctuations are then generated dynamically by the reaction currents. As a validation, we also initialize the unbiased chemical subblock with its equilibrium covariance \(T\chi/a^3\). At \(x_{\rm in}=100\) this changes \(n_s\) by at most \(2.1\times10^{-8}\) at the strong, intermediate, and weak validation points. Moving the initialization surface from \(x_{\rm in}=100\) to \(200\) changes \(n_s\) by at most \(7.6 \times 10^{-5}\) and the pivot amplitude by at most \(2.5\times10^{-3}\).

The Bunch-Davies prescription above corresponds to the \(n_*=0\) inflaton statistical state; it is not a claim that a thermal inflaton population is unphysical. The interplay of quantum and thermal fluctuation sources in WI has been studied more generally in stochastic formulations~\cite{Ramos:2013nsa}. 
Following Ref.~\cite{ORamos:2025uqs} we therefore solve a separate \(n_*=n_{\rm BE}\) sensitivity branch, with \(n_{\rm BE}\equiv \left[\exp\left(H_*/T_*\right)-1\right]^{-1}\). In that branch, the conventional thermal statistical inflaton source is included in the same enlarged covariance system, while the initial vacuum inflaton block is removed so that the same statistical population is not counted twice. Both statistical-state branches are independently amplitude normalized. The kinetic criterion used to assess when the thermal alternative is plausible is the hard-thermal-loop scattering ratio \(\Gamma_{\rm scat}/H\) of Ref.~\cite{ORamos:2025uqs}; it is distinct from both \(N_{\rm sph} \) and  \(\Gamma_{h}/H\).

The Gaussian Langevin description also requires many microscopic events within the coarse-graining volume and time. We define
\begin{equation}
    N_{\rm sph}\equiv \Gamma_{\rm sph}H^{-4}\,, \qquad N_{23}\equiv 2L_{23}H^{-4}\,.
\end{equation}
Both $N_{\rm sph}$ and $N_{23}$ are dimensionless event counts per Hubble four-volume. Values much larger than unity support a Gaussian central-limit description of reaction noise, whereas values of order unity signal that the two-point Langevin closure is no longer parametrically justified.

Here \(N_{\rm sph}\) estimates the expected number of topological transitions in a spacetime cell with spatial volume \(H^{-3}\) observed for one Hubble time \(H^{-1}\); \(N_{23}\) has the analogous interpretation for the effective perturbative channel. These quantities test the many-event assumption underlying Gaussian coarse-graining, not gluon, quark, or inflaton thermalization rates and their numerical values are reported in Sec.~\ref{sec:numerical}.

\paragraph{Spectral extraction.} For every one of the  numerical scan values of \(\lambda\) on each physical branch, we integrate seven independent modes at fixed physical initialization depth
\begin{equation}\label{eq:seven_offsets}
    \ln\left(k/k_*\right)\in \left\{-0.08,\, -0.04,\, -0.02,\, 0,\, 0.02,\, 0.04,\, 0.08\right\}\,.
\end{equation}
Equivalently, \(k_j=k_*e^{x_j}\) with the seven \(x_j\) above, and every mode is assigned its own initialization time \(N_{\rm in}(k_j)\) satisfying \(k_j/\left[a(N_{\rm in})\, H(N_{\rm in})\right]=100\). We compare quadratic and cubic local polynomial fits, centered finite differences at \(\left|\ln(k/k_*)\right|=0.02\) and \(0.04\), and local cubic-spline derivatives. We take the median of the mutually consistent estimates as the central tilt and use their full spread as a numerical differentiation error estimate. Over the complete numerical calculation the maximum estimator spread in \(n_s\) is \(2.21\times 10^{-6}\) for the gauge+quark branch and \(2.41\times 10^{-6}\) for the quark-only branch. We quote the running only where the same independent estimators show comparable agreement. With \(x\equiv \ln(k/k_*)\) we use
\begin{equation}
\ln\mathcal P_{\mathcal R}=a_0+a_1x+\frac12a_2x^2+\frac16a_3x^3+\cdots,
\end{equation}
so that
\begin{equation}
n_s-1=\left.\frac{d\ln\mathcal P_{\mathcal R}}{d\ln k}\right|_{k_*},
\qquad
\alpha_s^{\rm run}=\left.\frac{dn_s}{d\ln k}\right|_{k_*}.
\end{equation}
For each tensor helicity, the metric perturbation obeys schematically
\begin{equation}
    \ddot{h}_\lambda+3H\dot{h}_\lambda+\frac{k^2}{a^2}h_\lambda=\frac{2}{M_{\rm Pl}^2}\Pi_\lambda^{\rm TT}\,.
\end{equation}
The missing microscopic object in the present scalar transport calculation is the finite-temperature QCD transverse-traceless anisotropic stress \(\Pi_{\lambda}^{\rm TT}\), including its unequal-time correlator. It could add tensor power and, in a parity-odd plasma, need not populate the two helicities equally. Since it is not computed here, we report only
\begin{equation}\label{eq:tensor_numerator_new}
r_{\rm vac}\equiv\frac{\mathcal P_T^{\rm vac}(k_*)}{\mathcal P_{\mathcal R}(k_*)},
\qquad \mathcal P_T^{\rm vac}=\frac{2H_*^2}{\pi^2M_{\rm Pl}^2},
\end{equation}
where the subscript ``vac'' indicates that the numerator is the vacuum tensor spectrum. We do not identify \(r_{\rm vac}\) with the complete tensor-to-scalar ratio when gauge or thermal anisotropic stress has not been computed.
\section{Numerical results}
\label{sec:numerical}
We use the pivot \(k_*=0.05\, \text{Mpc}^{-1}\) and normalize each branch to \(A_s=2.105\times10^{-9}\)~\cite{Planck:2018jri}. The background survey covers \(10^{-21}\leq \lambda\leq 10^{-15}\). The background is evolved to the geometric end condition \(\epsilon_H=1\).

Throughout this section, ``quark-only'' means the sphaleron system with the five axial response densities retained but the hard gauge-helicity variable and \(J_{23}\) switched off. ``Gauge+quark'' means the same system with \(n_h\), the sphaleron-induced source for \(n_h\), and \(J_{23}\) included. The two branches are independently amplitude normalized. We define
\begin{equation}\label{eq:QgQq_explicit}
    Q_q\equiv \frac{\Upsilon_{q}}{3H_q}\,,\qquad Q_g\equiv \frac{\Upsilon_{\rm tot}}{3H_g}\,.
\end{equation}
The superscripts $q$ and $g$ used below distinguish the independently normalized quark-only and gauge+quark solutions. A star denotes evaluation at the CMB pivot. Thus $Q_*^q$ and $Q_*^g$, for example, need not be compared at identical background field values, because each branch is separately normalized to the observed scalar amplitude. This independent normalization is essential when interpreting shifts of $n_s$, $N_*$, and $r_{\rm vac}$.\\ For concise terminology below, we refer to the region with $Q_*\gtrsim1$ as the strong warm branch, $Q_*\sim\mathcal O(0.1-1)$ as the intermediate branch, and $Q_*\ll1$ as the weak branch.
\subsection{Rate, renormalization-scale, and inflaton-state sensitivity}
\begin{figure}[t!]
\centering
\includegraphics[width=.85\textwidth]{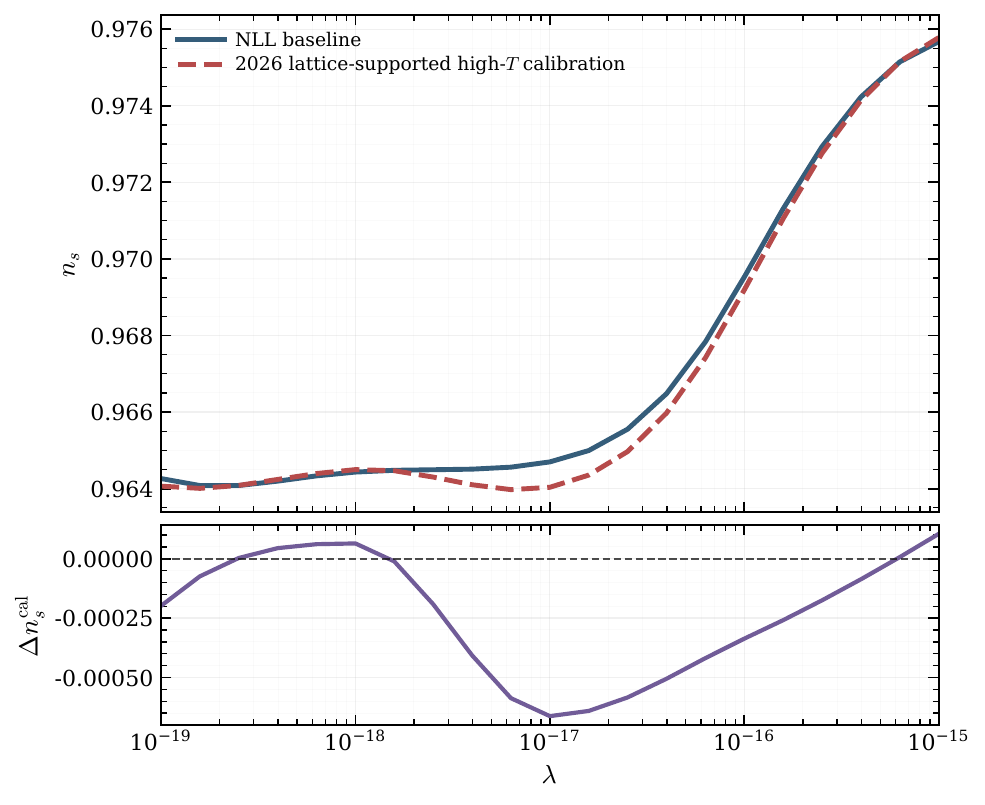}
\caption{\it
Scalar spectral index for the NLL sphaleron baseline and the high-temperature
$SU(3)$ calibration~\cite{Guin:2026kbp}. The upper panel shows the variation
of $n_s$ along the $\lambda$ scan, while the lower panel isolates the
calibration shift
$\Delta n_s^{\rm cal}\equiv n_s^{\rm high\text{-}T}-n_s^{\rm NLL}$
at fixed $\lambda$. The sub-$10^{-3}$ calibration effect should therefore
be distinguished from the total $\mathcal O(10^{-2})$ variation of $n_s$
along the inflationary trajectory.}
\label{fig:sph-calibration}
\end{figure}

The fiducial results use \(\mu_{\rm ren}=T\) for direct comparison with the baseline construction. We then repeat the background and scalar-spectrum calculation, including amplitude normalization, at the conventional thermal-QCD choices \(\mu_{\rm ren}=\pi T,\, 2\pi T,\, 4\pi T\). Across these independently solved benchmarks, the full \(\pi T\rightarrow 4\pi T\) variation changes the scalar tilt by at most \(1.86\times10^{-4}\) and the dissipation ratio \(Q_*\) by at most \(2.55\%\); the largest residual mismatch in the scalar amplitude after direct mode integration is \(2.37\times10^{-3}\), and the largest spread among the independent \(n_s\) estimator is \(2.49 \times10^{-6}\). These variations are smaller than the gauge-response displacement emphasized below.

The high-temperature sphaleron calibration changes \(n_s\) by at most \(6.63\times 10^{-4}\) at fixed \(\lambda\) over the directly recomputed benchmark points. The corresponding background shift in \(Q_*\) is largest at the strong end, while the tilt displacement remains smaller than the uncertainty associated with transport beyond linear response. Fig.~\ref{fig:sph-calibration} shows that the sensitivity to the sphaleron-rate normalization remains smaller than the shift produced by the gauge-helicity response over the numerical range.
The comparison is informative because $\Gamma_{\rm sph}$ enters both the bare dissipation and the screening parameters. The small vertical displacement between the two rate calibrations shows that the qualitative gauge-helicity effect is not produced by choosing one particular high-temperature normalization of the sphaleron rate. At the same time, the visible variation of $n_s$ along the $\lambda$ direction is much larger than the calibration shift, so the figure should be read as a rate-systematics test rather than as an uncertainty band on the full trajectory.

Because the coefficient $c_{\rm ch}$ in Eq.~\eqref{eq:Gamma_ch_f} is phenomenological, we also test the quark-screening baseline against the two-decade variation $c_{\rm ch}=10^{-3},\,10^{-2},\,10^{-1}$. Holding the fiducial gauge+quark background points fixed and recomputing the flavor-resolved factor $D_q$, the largest departure from the $c_{\rm ch}=10^{-2}$ result over $10^{-21}\leq\lambda\leq10^{-15}$ is $1.68\%$, occurring at the smallest-$\lambda$ end; for $\lambda\geq10^{-19}$ the deviation is below $0.44\%$ and decreases rapidly toward the weak branch. The quark-screening baseline is therefore nearly unchanged across this range, so the additional screening associated with the hard-helicity response cannot be attributed to the fiducial choice of $c_{\rm ch}$. This fixed-background test is not a substitute for a first-principles finite-temperature determination of the chirality-changing rates, but it shows that the structural gauge-helicity effect is insensitive to an order-of-magnitude variation of the phenomenological prefactor in either direction.
\begin{figure}[t!]
\centering
\includegraphics[width=.91\textwidth]{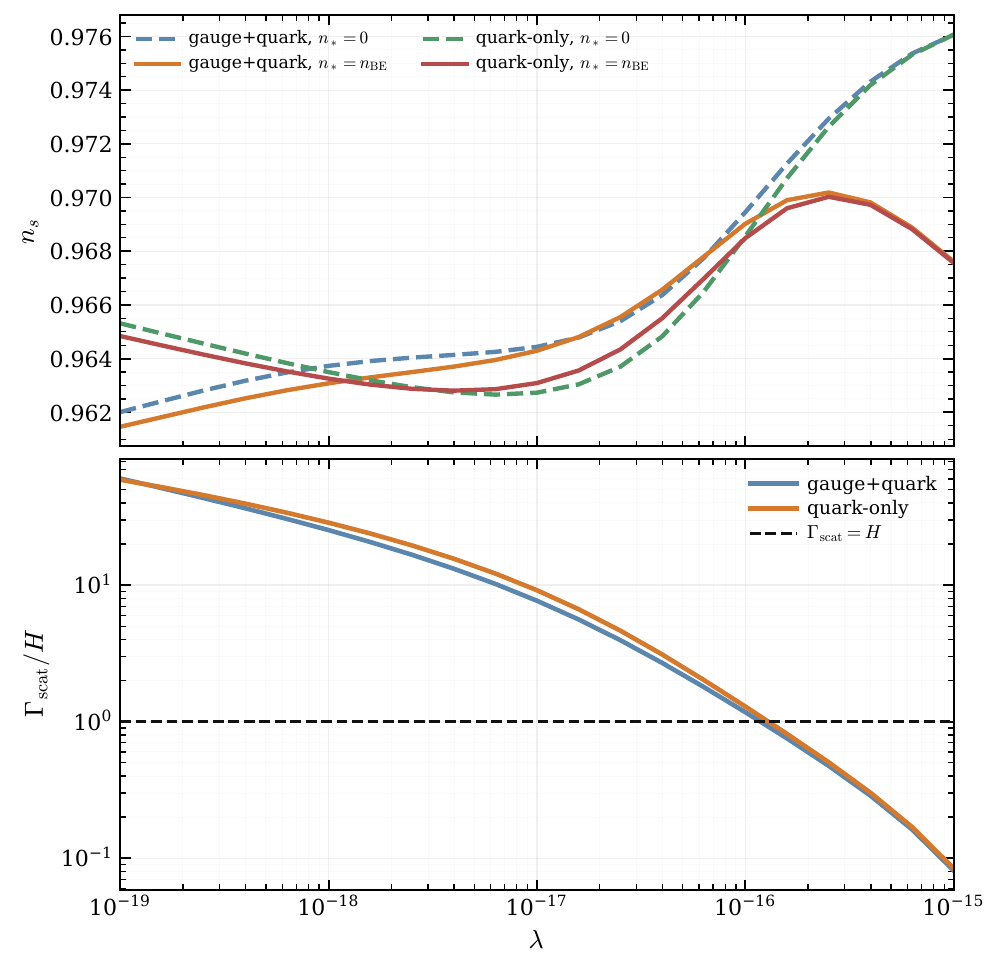}
\caption{\it  Sensitivity to the inflaton statistical state after independent amplitude normalization. The upper panel compares the $n_*=0$ and $n_*=n_{\rm BE}$ scalar tilts; the lower panel shows the hard-thermal-loop scattering ratio $\Gamma_{\rm scat}/H$. The thermal branch is interpreted as physically motivated where the scattering rate exceeds the Hubble rate and elsewhere it is shown only to illustrate the dependence on the assumed inflaton statistical state. The horizontal reference $\Gamma_{\rm scat}/H=1$ marks where microscopic
inflaton scattering becomes comparable to the Hubble expansion rate.}
\label{fig:thermal_inflaton}
\end{figure}

\begin{figure}[t!]
\centering
\includegraphics[width=.88\textwidth]{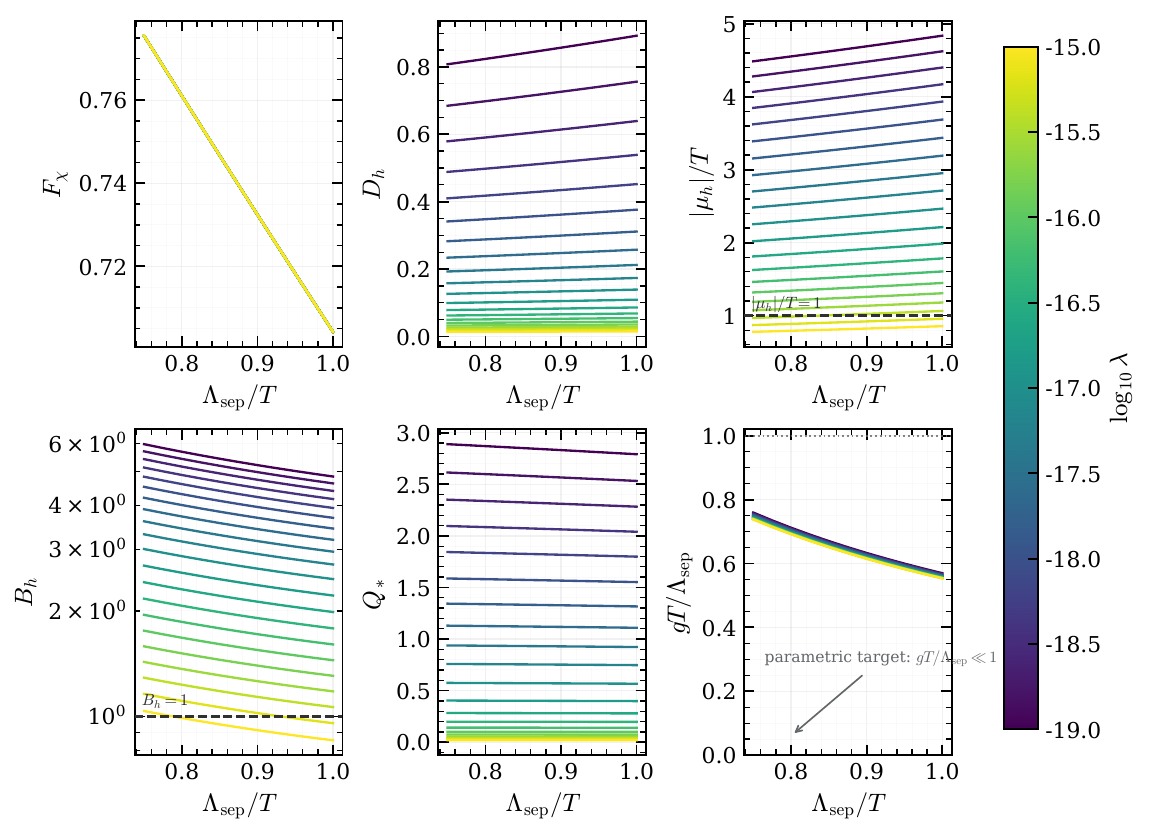}
\caption{\it  Hard-mode matching checks over the numerical interval
$10^{-19}\leq\lambda\leq10^{-15}$.
The separation scale is restricted to
$0.75\leq\Lambda_{\rm sep}/T\leq1$.
The six panels show $F_\chi$, $D_h$, $|\mu_h|/T$, $B_h$, $Q_*$,
and $gT/\Lambda_{\rm sep}$, with a common $\lambda$ color scale
throughout. The dashed references $|\mu_h|/T=1$ and $B_h=1$
mark, respectively, the onset of order-one gauge bias and the
boundary of a literal finite-density Bose interpretation.
The lower-right panel indicates the parametric target
$gT/\Lambda_{\rm sep}\ll1$. Significant gauge screening can
therefore coexist with a hard-soft hierarchy that is not
parametrically broad.}
\label{fig:cutoff_final}
\end{figure}

The independently amplitude-normalized thermal gauge+quark benchmarks \((Q_*,n_s)\) at \(\lambda=10^{-19},\, 10^{-18},\, 10^{-17},\, 10^{-16},\, 10^{-15}\) are $(3.037,0.9615)$, $(1.562,0.9631)$, $(0.523,0.9643)$, $(0.0830,0.9690)$, and $(0.00587,0.9676)$. The corresponding scattering ratios are $59.1$, $28.7$, $9.17$, $1.29$, and $0.084$. Thus thermal inflaton occupation is kinetically well motivated on the strong branch and over much of the intermediate branch, but not at the weakest endpoint. We therefore retain \(n_*=0\) as the fiducial choice, and present \(n_*=n_{\rm BE}\) as a separately normalized test of the dependence on the inflaton statistical state. Across all the normalized thermal solutions, the largest residual mismatch in the pivot amplitude is \(2.60\times10^{-3}\), and the largest spread among the \(n_s\) estimator is \(1.46\times10^{-6}\). Fig.~\ref{fig:thermal_inflaton} makes clear that thermal inflaton occupation is a dynamical regime choice rather than a universal correction: it is supported where \(\Gamma_{\rm scat}/H\gtrsim 1\) and loses kinetic justification on the weak branch. The two panels also separate statistical-state sensitivity from background thermalization. Where $\Gamma_{\rm scat}/H\gg1$, the thermal occupation can alter the tilt at a visible level after amplitude renormalization; as the ratio drops below unity, that branch ceases to have a kinetic justification even if it remains numerically well defined. This prevents the thermal-vacuum difference from being interpreted as a uniform theory error across the scan.
\subsection{Background response and transport checks}
The formal hard-soft hierarchy \(g^2 T\ll gT\ll \Lambda_{\rm sep}\lesssim T\) is also checked numerically rather than assumed. Over the numerical parameter range the gauge coupling is \(g\simeq 0.554-0.570\). Restricting the hard-mode consistency check to \(0.75\leq \Lambda_{\rm sep}/T\leq 1\) gives \(0.55 \lesssim gT/\Lambda_{\rm sep}\lesssim 0.76\) and \(0.31\lesssim g^2T/\Lambda_{\rm sep}\lesssim 0.43\). The available interval therefore does not realize a parametrically wide factorization window. Fig.~\ref{fig:cutoff_final} is therefore a local test of the hard-soft matching, not a controlled factorization-scale uncertainty band; values \(\Lambda_{\rm sep}>T\) are excluded from the literal hard-mode definition. Across this restricted scan \(0.855\lesssim B_h\lesssim 5.98\), so the Bose-positivity boundary is crossed within the phenomenologically relevant region.
\begin{figure}[t!]
\centering
\includegraphics[width=.96\textwidth]{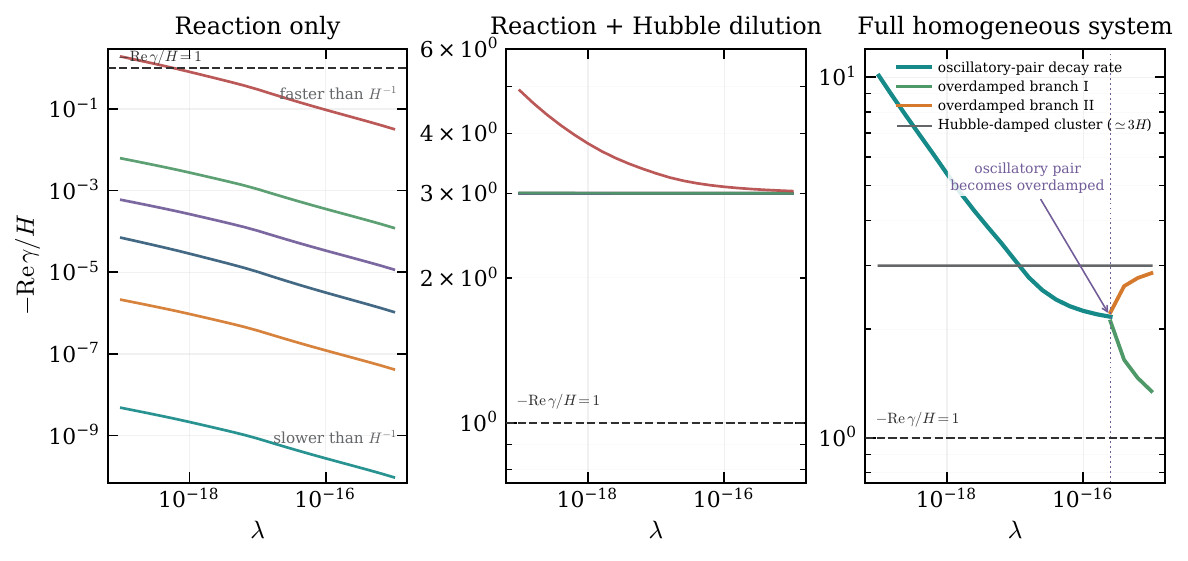}
\caption{\it  Relaxation spectrum over
$10^{-19}\leq\lambda\leq10^{-15}$.
The panels show the reaction-only spectrum, the reaction spectrum
including Hubble dilution, and the stable modes of the full
homogeneous system. The horizontal reference
$-\mathrm{Re}\,\gamma/H=1$ corresponds to relaxation in one Hubble
time: modes below it are slower than $H^{-1}$, whereas modes above
it decay within a Hubble time. Hubble dilution substantially
reorganizes the damping hierarchy even though several microscopic
reaction directions are not parametrically fast. In the full
system, the apparent two-branch structure at large $\lambda$
occurs when a complex-conjugate oscillatory pair becomes overdamped
and separates into two real decay rates.}
\label{fig:relaxation}
\end{figure}

\begin{figure}[t!]
\centering
\includegraphics[width=.92\textwidth]{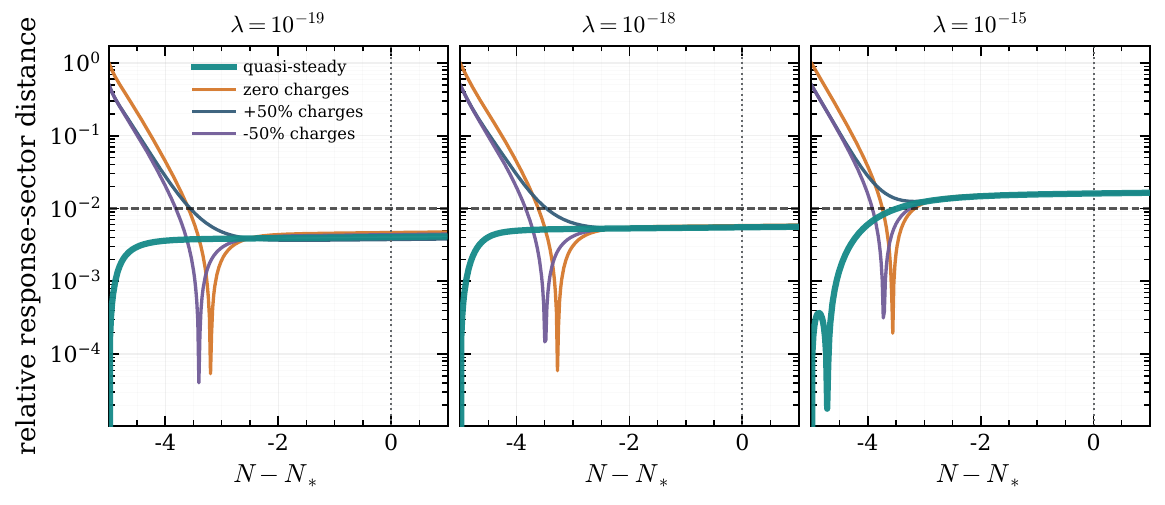}
\caption{\it  Attractor test for representative strong, intermediate, and weak points. The full homogeneous system is initialized on the quasi-steady response manifold, with vanishing response densities, and with $\pm50\%$ displacements. The comparison tests the quasi-steady background directly rather than inferring its accuracy from the relaxation rates alone.}
\label{fig:full-vs-reduced}
\end{figure}

For each slowly relaxing quark flavor we retain
\begin{equation}
    n_{5f}=\chi_q\mu_{5f}\,,\qquad\chi_q=\frac{N_cT^2}{3}\,,
\end{equation}
and \(\Gamma_{{\rm ch},f}\) defined in Eq.~\eqref{eq:Gamma_ch_f}. The coupled reaction network is
\begin{align}
    \dot{n}_{5f}+3Hn_{5f}&=-2J_{\rm sph}-\Gamma_{{\rm ch},f}\,n_{5f}\,,\label{eq:n_dot_5f_def}\\
    \dot{n}_h+3Hn_h&=\frac{2\zeta_h}{\alpha_s}J_{\rm sph}+J_{23}\,.\label{eq:n_dot_h_def}
\end{align}
\begin{figure}[t!]
\centering
\includegraphics[width=.85\textwidth]{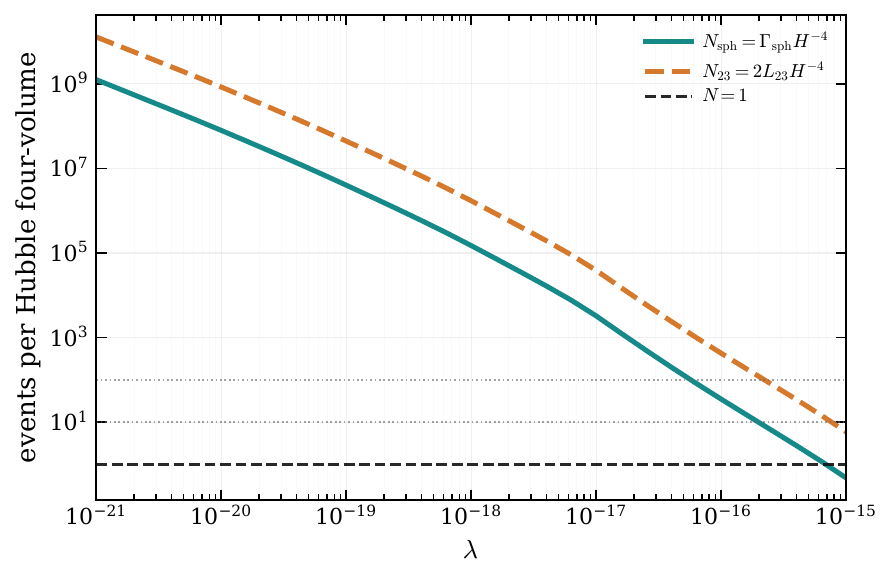}
\caption{\it
Reaction counts per Hubble spacetime cell.
$N_{\rm sph}=\Gamma_{\rm sph}H^{-4}$ and
$N_{23}=2L_{23}H^{-4}$ diagnose whether many-event Gaussian coarse
graining is plausible; they are not plasma thermalization rates.
The horizontal reference $N=1$ marks the boundary below which even the
minimal many-event criterion fails, while parametrically controlled
Gaussian coarse graining requires $N\gg1$.}\label{fig:event-counts}
\end{figure}

\begin{table*}[t!]
\scriptsize
\centering
\setlength{\tabcolsep}{1.8pt}
\begin{tabular}{ccccccccccccc}
\hline
$\lambda$
& $Q_q$
& $Q_g$
& $N_*^q$
& $N_*^g$
& $n_s^q$
& $n_s^g$
& $r_{\rm vac}^g$
& $D_h$
& $|\mu_h|/T$
& $|A_{23}|$
& $|A_{\rm sph}|$
& $\xi_{\rm CS}$\\
\hline
$10^{-19}$
& 5.14
& 3.13
& 54.9
& 54.9
& 0.9653
& 0.9620
& $4.13\times10^{-8}$
& 0.621
& 3.66
& 5.91
& 226
& 371\\

$10^{-18}$
& 2.48
& 1.67
& 55.9
& 55.8
& 0.9634
& 0.9637
& $8.74\times10^{-7}$
& 0.264
& 2.70
& 10.2
& 403
& 512\\

$10^{-17}$
& 0.80
& 0.59
& 56.9
& 56.9
& 0.9627
& 0.9644
& $2.27\times10^{-5}$
& 0.098
& 1.77
& 18.0
& 724
& 797\\

$10^{-16}$
& 0.11
& 0.10
& 57.8
& 57.9
& 0.9685
& 0.9695
& $4.92\times10^{-4}$
& 0.031
& 1.03
& 33.0
& $1.35\times10^3$
& $1.39\times10^3$\\

$10^{-15}$
& 0.019
& 0.018
& 58.6
& 58.7
& 0.9760
& 0.9761
& $6.32\times10^{-3}$
& 0.011
& 0.60
& 57.3
& $2.35\times10^3$
& $2.37\times10^3$\\
\hline
\end{tabular}
\caption{\it  Numerical benchmarks for the independently amplitude-normalized quark-only and gauge+quark branches.
The dissipation ratios $Q_q$ and $Q_g$ are defined in Eq.~\eqref{eq:QgQq_explicit}.
The quantities $N_*^q$ and $N_*^g$ give the pivot-to-end e-fold numbers for the independently normalized quark-only and gauge+quark backgrounds. The slow-roll relation is summarized in Eq.~\eqref{eq:N_*_simeq_for_quartic}; in the numerical calculation, however, $Q(\phi)$ is evolved rather than treated as constant.
Every quoted value of $n_s$ is obtained from the uniform numerical scan covariance calculation of Eq.~\eqref{eq:seven_offsets}.
The quantity $D_h$ measures the strength of gauge screening; $|\mu_h|/T$, $|A_{23}|$, and $|A_{\rm sph}|$ diagnose departures from the small-bias regime, while $\xi_{\rm CS}$ measures the unscreened topological drive.
The quantity $r_{\rm vac}$ uses the vacuum tensor spectrum defined in Eq.~\eqref{eq:tensor_numerator_new}.}
\label{tab:benchmarks} 
\end{table*}

The factor \(\zeta_h\) is kept explicit because converting one topological transition into the effective hard-helicity variable requires matching the collective soft configuration to the kinetic description. The \(1/\alpha_s\) scaling follows the parametric many-soft-quanta character of a non-Abelian topological transition discussed in Ref.~\cite{Broadberry:2025ggb}; \(\zeta_h\) encodes the order-unity matching that cannot be fixed by this parametric argument alone.

Fig.~\ref{fig:cutoff_final} makes the hard-soft limitation quantitative: the matched hard-helicity variable remains useful, but the available scale hierarchy is not broad enough to define a controlled factorization uncertainty. Taken together, the panels show that increasing $\Lambda_{\rm sep}/T$ changes the free hard susceptibility through $F_\chi$ and correspondingly changes the inferred $D_h$ and $B_h$, while $gT/\Lambda_{\rm sep}$ remains of order unity rather than becoming asymptotically small. The simultaneous appearance of appreciable $D_h$ and $|\mu_h|/T\gtrsim1$ explains why the screening effect can be numerically important precisely where a literal finite-density Bose interpretation is least controlled.
\begin{figure}[t!]
\centering
\includegraphics[width=.92\textwidth]{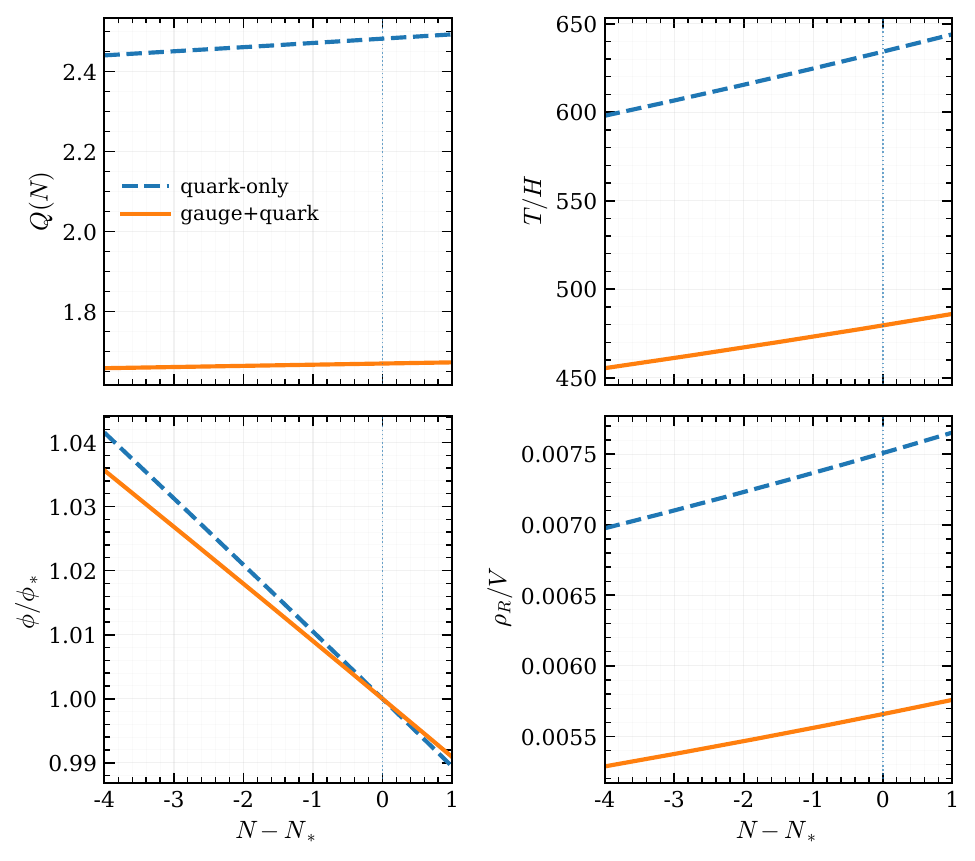}
\caption{\it  Direct background comparison at $\lambda=10^{-18}$ after independent amplitude normalization. The quark-only and gauge+quark branches are shown through $Q(N)$, $T/H$, $\phi/\phi_*$, and $\rho_R/V$ around the pivot. This shows directly how the hard-helicity response changes the background, rather than inferring the effect only from $n_s$.}
\label{fig:background_branch_compare}
\end{figure}

The relaxation spectrum confirms the conceptual point of Eq.~\eqref{eq:quasiadiabatic_approx}: several reaction directions are not parametrically fast, and Hubble dilution supplies an important part of their damping. In the fiducial scan \(R_h=\Gamma_h/(3H)\) remains between approximately \(3.9\times10^{-5}\) and \(1.2\times10^{-2}\) even though \(D_h\) can reach order unity. It is therefore inconsistent either to assume instantaneous hard-helicity equilibration or to set \(\mu_h=0\). Fig.~\ref{fig:relaxation} therefore shows that the gauge-helicity response cannot be eliminated on the basis of a parametrically fast microscopic relaxation rate. The reaction-only panel tests microscopic relaxation, whereas adding $3H$ shows how cosmological dilution damps otherwise slow charge directions. The full homogeneous eigenvalues therefore need not be microscopically large for the background to approach a local attractor. This also explains why there is no contradiction with Fig.~\ref{fig:full-vs-reduced}: fast algebraic elimination is not parametrically justified, yet the reduced homogeneous trajectory can still be numerically accurate near the pivot because Hubble damping and source feedback stabilize the background.
\begin{figure}[t!]
\centering
\includegraphics[width=.75\textwidth]{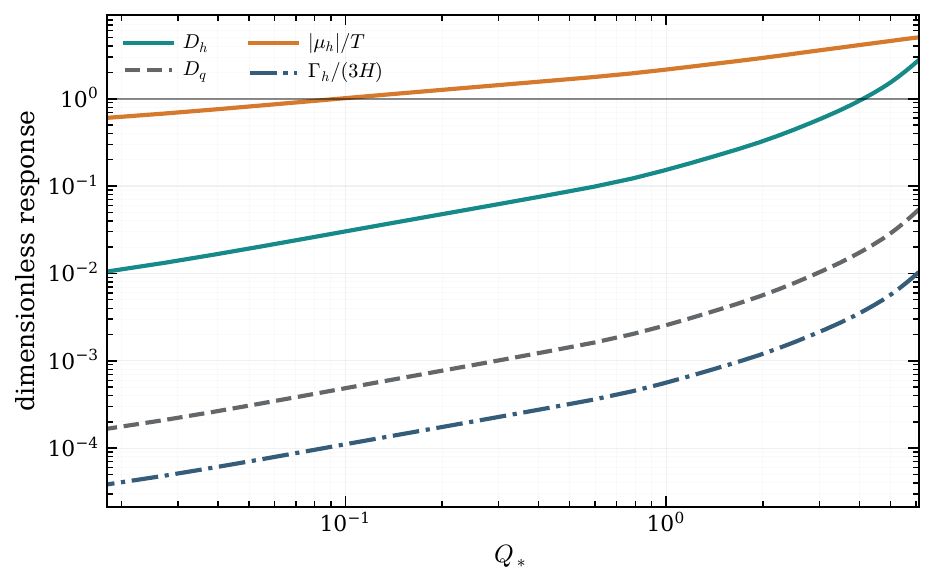}
\caption{\it  Hierarchy of response and relaxation scales in the amplitude-normalized gauge+quark calculation over $10^{-21}\leq\lambda\leq10^{-15}$. The quantities $D_h$ and $D_q$ measure gauge-helicity and quark screening, $|\mu_h|/T$ measures the induced chemical bias, and $\Gamma_h/(3H)$ compares perturbative hard-helicity relaxation with expansion.}
\label{fig:response}
\end{figure}

\begin{figure}[t!]
\centering
\includegraphics[width=.75\textwidth]{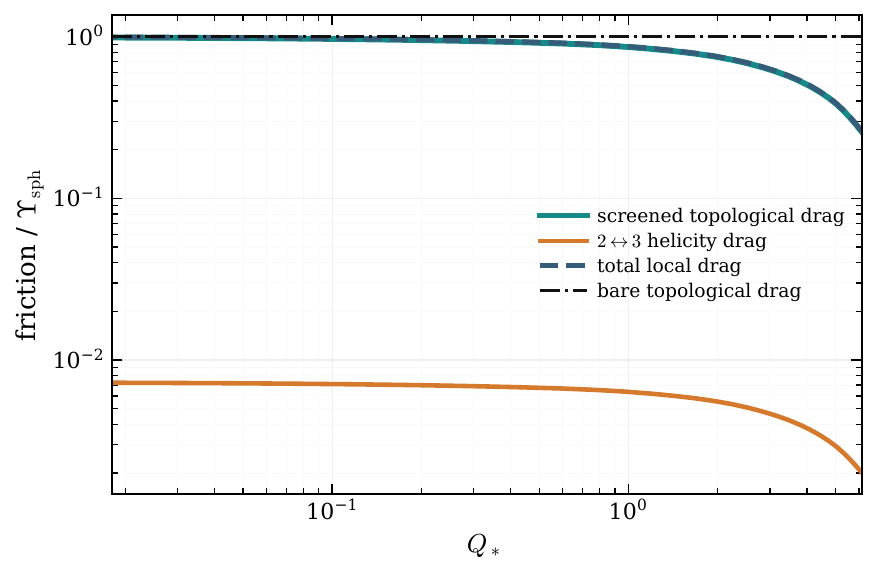}
\caption{\it  Contributions to the effective dissipation in units of the zero-density sphaleron friction. Gauge-helicity screening suppresses the topological contribution at large $Q$, while the direct dissipative contribution from the effective $2\leftrightarrow3$ channel remains below the percent level for the fiducial matching.}
\label{fig:drag_two}
\end{figure}

Fig.~\ref{fig:full-vs-reduced} nevertheless shows that the quasi-steady background is a local attractor near the pivot, separating its numerical accuracy from the absence of a parametric time-scale hierarchy. The main effect of the gauge response is a reduction of the warm dissipation ratio. At \(\lambda=10^{-19}\), for example, \(Q_*\) changes from \(5.14\) on the quark-only branch to \(3.13\) on the gauge+quark branch. The difference decreases continuously toward weak dissipation. This behavior tracks \(D_h\): the hard-helicity response screens a substantial fraction of the topological driving term at strong and intermediate coupling, while \(D_h\ll1\) on the weak branch. The numerical reduction agrees with the analytic expectation from Eq.~\eqref{eq:simplieifed_total_affinity_R_h_def}: at fixed quark screening, increasing $D_h$ lowers $S/\dot\phi$ and therefore lowers the sphaleron part of $\Upsilon_{\rm tot}$. The attractor test adds information that the algebraic formula alone cannot provide: trajectories started off the quasi-steady charge manifold converge back toward the same local background, showing that the observed screening is not an artifact of initializing directly on the reduced solution.
\begin{figure}[t!]
\centering
\includegraphics[width=.75\textwidth]{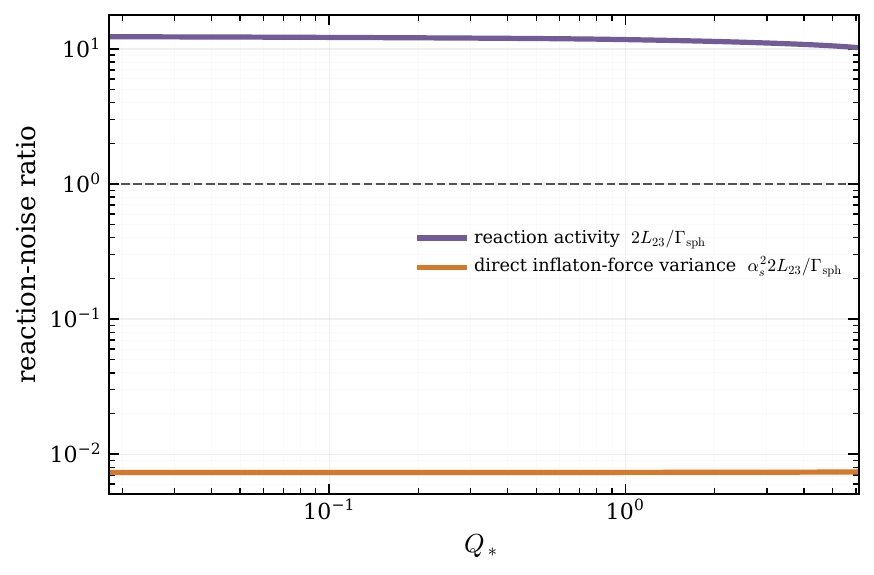}
\caption{\it  Relative stochastic forcing along the exploratory scan. Although the perturbative helicity-changing channel has a large reaction-rate coefficient, its projection onto the inflaton force carries additional powers of $\alpha_s$, leaving the direct scalar-noise contribution below the percent level relative to the sphaleron contribution.}
\label{fig:noise_updated}
\end{figure}

Fig.~\ref{fig:event-counts} thus tests the many-event assumption underlying the Gaussian Langevin description, independently of any condition for thermal equilibration. At the weakest endpoint, \(\lambda=10^{-15}\), the pivot value falls to \(N_{\rm sph}\simeq 0.49<1\), so the Hubble-volume many-event criterion underlying Gaussian sphaleron noise is not satisfied there. We therefore retain this point as a boundary diagnostic, rather than as a parametrically controlled stochastic prediction. The effective \(2\leftrightarrow3\) channel contributes only about \(0.2\%-0.7\%\) of the zero-density sphaleron friction directly. Its principal role is instead to determine the relaxation of the hard-helicity density and hence the self-consistent value of \(\mu_h\). The same hierarchy appears in the stochastic sector. Fig.~\ref{fig:background_branch_compare} shows that the reduction in \(Q\) is already a property of the background dynamics and is not generated by the subsequent extraction of the scalar spectrum.
Fig.~\ref{fig:background_branch_compare} also clarifies why changes in observables need not scale directly with the instantaneous change in $Q_*$. The altered dissipation changes $T/H$, the field trajectory, and $\rho_R/V$ coherently over several $e$-folds around the pivot. The scalar spectrum therefore responds to an integrated change of the coupled background and perturbation system, rather than to a single local value of the friction coefficient.

\begin{figure}[t!]
\centering
\includegraphics[width=.96\textwidth]{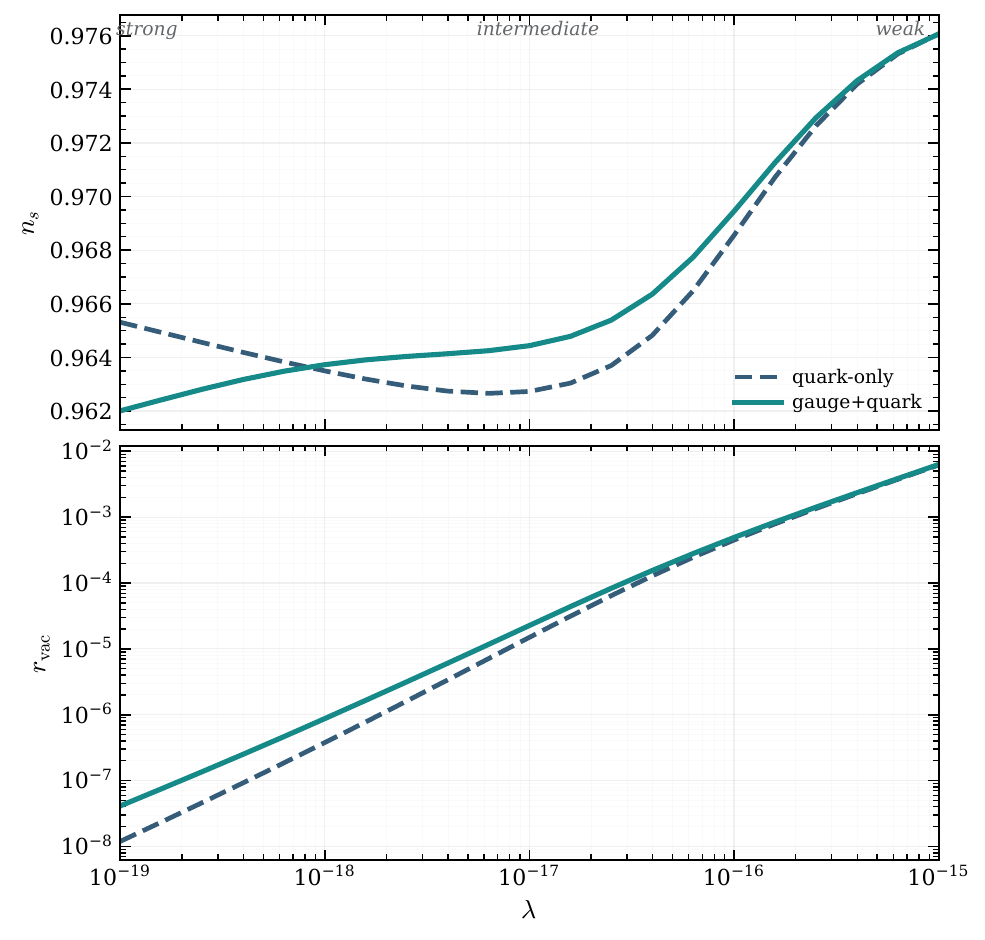}
\caption{\it
Uniform numerical scan over $10^{-19}\leq\lambda\leq10^{-15}$.
The upper panel shows the scalar spectral index and the lower panel
$r_{\rm vac}$, defined using the vacuum tensor spectrum, for the
quark-only and gauge+quark branches. The strong, intermediate, and weak
labels indicate the qualitative dissipation regimes used throughout the
text and should not be interpreted as sharp boundaries.}
\label{fig:spectra_new}
\end{figure}

The complete numerical trajectory is depicted in Fig.~\ref{fig:spectra_new}. At the five benchmarks level, Tab.~\ref{tab:benchmarks} shows that the gauge-induced shifts in \(n_s\) are \(-3.30\times10^{-3},\, 2.32\times10^{-4},\, 1.71\times10^{-3},\, 8.95\times10^{-4}\), and \(-4.21\times10^{-6}\) from \(\lambda=10^{-19}\) to \(10^{-15}\). The corresponding ratios \(r_{\rm vac}^g/r_{\rm vac}^q\) are \(3.46,\, 2.31,\, 1.51,\, 1.095,\) and \(1.022\). The altered background can therefore change \(r_{\rm vac}\) appreciably even where the shift in the scalar tilt is modest.

Fig.~\ref{fig:response} isolates the underlying mechanism: substantial screening can coexist with \(\Gamma_h/H\ll1\), so a slowly relaxing gauge response can still modify the mean topological drive. Fig.~\ref{fig:drag_two} shows this effect is primarily indirect: gauge helicity suppresses the sphaleron contribution through screening, while the direct \(2\leftrightarrow3\) drag remains small.
Analytically, this is already encoded in the different structures of $D_h$ and $R_h$: $D_h$ is proportional to the sphaleron diffusion rate divided by the helicity susceptibility and total relaxation scale, whereas $R_h$ contains only the perturbative helicity relaxation relative to expansion. Numerically, Fig.~\ref{fig:response} confirms that these quantities are not interchangeable. Fig.~\ref{fig:drag_two} then verifies the consequence directly in the force decomposition: the important effect is the suppression of the sphaleron channel, not the addition of a comparably large $2\leftrightarrow3$ friction term.
\begin{figure}[t!]
\centering
\includegraphics[width=.93\textwidth]{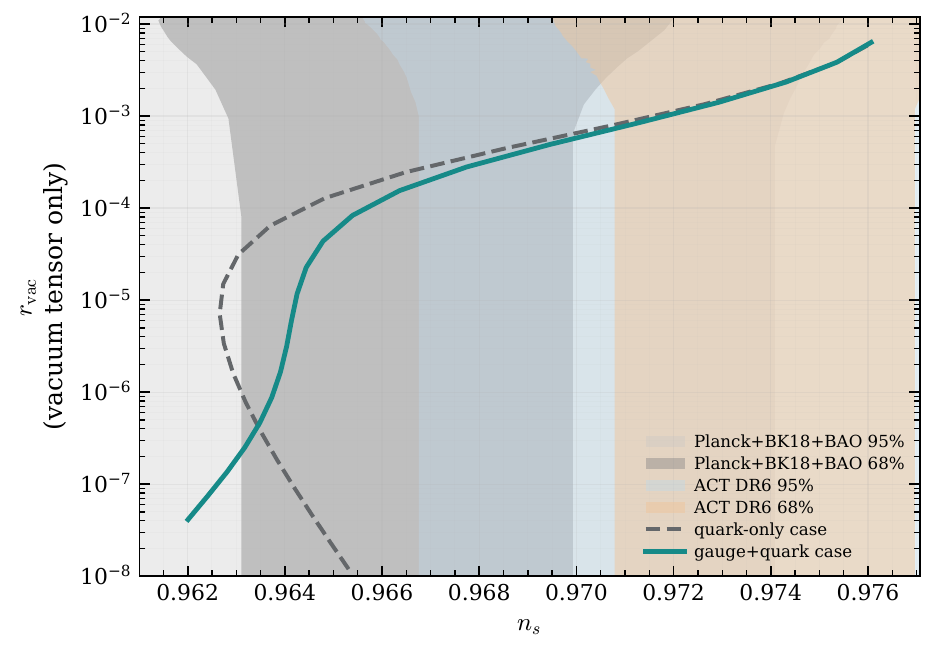}
\caption{\it  Numerical trajectories in the
$(n_s,r_{\rm vac})$ plane.
The theoretical curves display
$r_{\rm vac}$,
containing only the vacuum tensor contribution of
Eq.~\eqref{eq:tensor_numerator_new}; this restriction is stated
explicitly on the vertical axis. The shaded
\emph{Planck}+BK18+BAO~\cite{BICEP:2021xfz} and ACT DR6~\cite{AtacamaCosmologyTelescope:2025nti} contours constrain the total
tensor-to-scalar ratio $r$ and are shown \textbf{only} for orientation.
The overlay is therefore measure rather than a complete
likelihood comparison until the finite-temperature
transverse-traceless stress contribution is known.}
\label{fig:nsr_new}
\end{figure}
\paragraph{Validity hierarchy.} For clarity, we distinguish three levels of statements in what follows. Relations obtained from the specified reaction network and linear-response closure are model derivations; convergence, attractor behavior, and observable shifts are numerical results within that closure; locality, Gaussian finite-affinity noise, the one-moment hard-helicity closure, and the matched hard-channel coefficients remain microscopic assumptions.

 The numerical trajectories satisfy the inflationary and operator-level hierarchies while probing a different limitation in the transport sector. Over the broader background scan, \(T/H>1,\, \Lambda_{\rm EFT}/T\gg 1,\) and \(V/\rho_R\gg1\). These conditions establish a warm inflationary background below the EFT scale inferred from the dimension-five operator, but they do not imply small affinity, a broad hard-soft factorization window, or a short microscopic memory time. Fig.~\ref{fig:noise_updated} shows the analogous hierarchy in the stochastic sector: the \(2\leftrightarrow3\) channel is important for determining \(\mu_h\) but remains a subdominant direct source of scalar noise.
It is useful here to distinguish reaction activity from direct scalar forcing. Although the hard-helicity channel can have sufficient event activity to regulate the response density, its projection onto the inflaton equation carries additional powers of $\alpha_s$. The noise plot therefore mirrors the drag decomposition: the channel can be dynamically important through the state it relaxes while remaining small as a direct source term for $\delta\phi$.
\begin{figure}[t!]
\centering
\includegraphics[width=.85\textwidth]{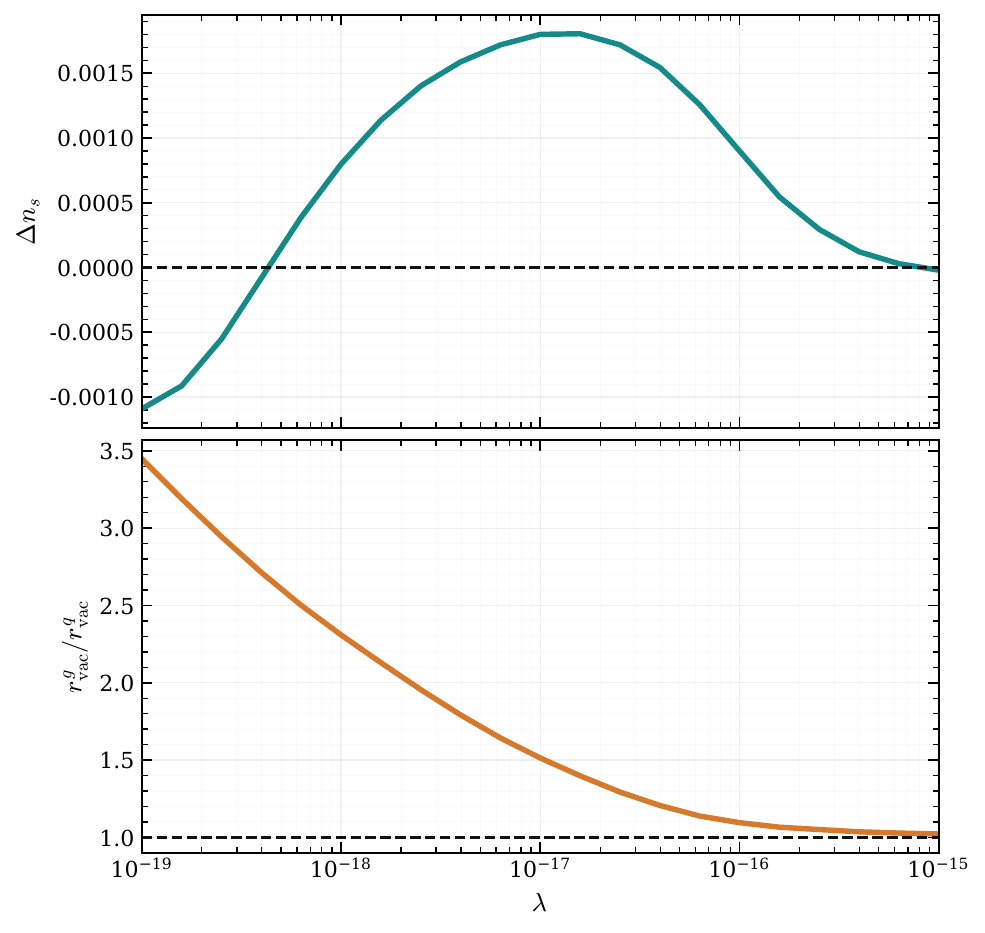}
\caption{\it  Changes induced by the gauge-helicity response across the uniform numerical scan. The upper panel shows $\Delta n_s=n_s^g-n_s^q$ and the lower panel shows $r_{\rm vac}^g/r_{\rm vac}^q$. The horizontal references $\Delta n_s=0$ and
$r_{\rm vac}^{g}/r_{\rm vac}^{q}=1$ denote no gauge-induced displacement.}
\label{fig:shift_new}
\end{figure}

\begin{figure}[t!]
\centering
\includegraphics[width=.85\textwidth]{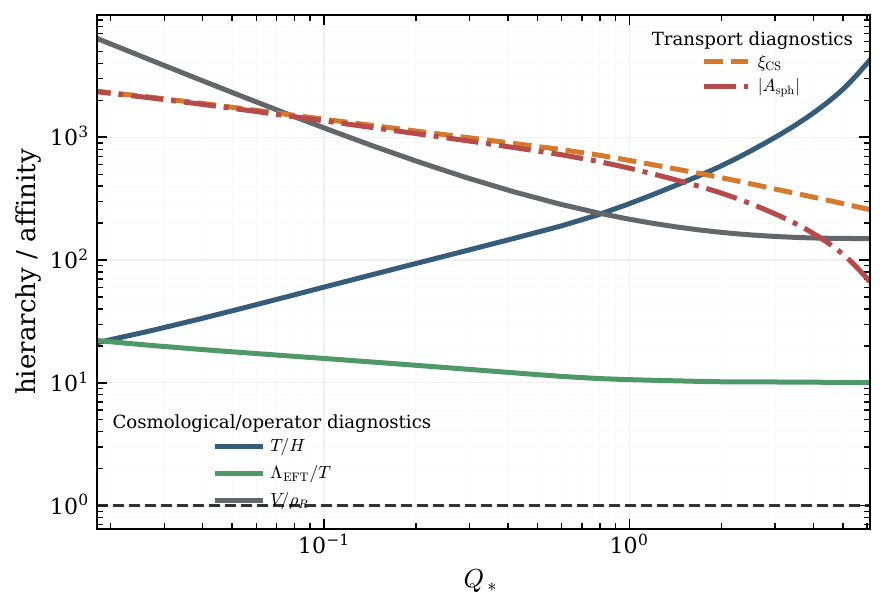}
\caption{\it
Hierarchy of physical scales along the gauge+quark background scan.
Solid curves collect the cosmological and operator-level diagnostics
$T/H$, $\Lambda_{\rm EFT}/T$, and $V/\rho_R$, while the dashed families
show the transport diagnostics $\xi_{\rm CS}$ and $|A_{\rm sph}|$.
The horizontal line marks the common order-one reference. The warm
condition, operator-level cutoff hierarchy, and inflationary energy
hierarchy remain separated from unity, whereas the unscreened topological
bias and sphaleron affinity are large. Parametric linear-response control
would require the relevant affinities to be much smaller than unity.}
\label{fig:validity_new}
\end{figure}

The agreement among the independent polynomial, finite-difference, and spline estimators, with a maximum spread in \(n_s\) below \(2.5\times10^{-6}\) over the full numerical calculation, verifies that the extracted tilt is insensitive to the choice of local spectral estimator. The full amplitude-normalized trajectories in Fig.~\ref{fig:spectra_new} further show that the gauge-helicity displacement persists across the numerical range rather than arising from a few selected benchmark points. In the $(n_s,r_{\rm vac})$ plane, Fig.~\ref{fig:nsr_new} places the two branches in observable space, with \(r_{\rm vac}\) defined using only the vacuum tensor contribution of Eq.~\eqref{eq:tensor_numerator_new}. Fig.~\ref{fig:shift_new} then makes clear that the scalar and tensor responses need not track one another: on the intermediate branch, a small shift in \(n_s\) can accompany an order-unity change in \(r_{\rm vac}\). The hierarchy in Fig.~\ref{fig:validity_new} places these results in their effective-theory context. The loss of small-affinity control in the transport sector is distinct from a breakdown of the inflationary derivative expansion, while \(T/H\gg1\) alone does not justify integrating out a response variable that remains dynamically slow. We therefore retain the response variables explicitly in the perturbation system, even though the homogeneous charge background admits an accurate quasi-steady reduction over much of the numerical parameter range.

Fig.~\ref{fig:spectra_new} provides the most direct comparison of the two amplitude-normalized predictions. The gauge and quark-only curves separate smoothly rather than at isolated points, confirming that the benchmark shifts in Tab.~\ref{tab:benchmarks} are samples of a coherent trajectory. The non-monotonic sign of $\Delta n_s$ is also physically informative: gauge screening changes both the background dissipation history and the coupled fluctuation transfer, so the tilt is not expected to follow a simple monotonic function of $Q_*$. By contrast, the ratio $r_{\rm vac}^g/r_{\rm vac}^q$ can remain substantially above unity because independent amplitude normalization changes the Hubble scale entering the vacuum tensor numerator while the scalar denominator is fixed at the pivot; this differing scalar and tensor response is explicit in Figure~\ref{fig:shift_new}.
\section{Discussion and conclusions}
\label{sec:discussion}
The central result of this work is that gauge backreaction changes the transport content of SMWI. Once a gauge-helicity response is induced by the rolling $\phi G\widetilde G$ interaction and does not relax parametrically faster than the cosmological modes of interest, it cannot consistently be represented only through a screened scalar friction coefficient. The response must remain dynamical. Our enlarged system implements this requirement by evolving the quark axial densities and the matched hard-helicity density together with the inflaton, radiation, and their correlated reaction noise.

Fig.~\ref{fig:validity_new} summarizes the central hierarchy of the paper in one place. The large values of $T/H$, $\Lambda_{\rm EFT}/T$, and $V/\rho_R$ show that the cosmological warm regime, the operator-level scale separation, and potential domination can coexist. The simultaneous large values of $\xi_{\rm CS}$ and $|A_{\rm sph}|$ demonstrate that these successful cosmological hierarchies do not imply proximity to the unbiased plasma state required for parametric linear-response control. This is why the principal limitation of the calculation is microscopic transport rather than slow roll or the dimension-five operator expansion.

Within this transport theory, gauge helicity has a substantial and physically transparent effect. The dominant modification does not come from a large additional drag. Instead, the induced helicity feeds back on the sphaleron affinity and screens the topological current. On the strong and intermediate branches this reduces the warm dissipation ratio and shifts the independently amplitude-normalized primordial trajectory. The effect can be large in the background even when the direct $2\leftrightarrow3$ force and its direct scalar-noise projection remain below the percent level. This separation between indirect screening and direct dissipation is central to the result: a slowly relaxing response can reorganize the mean anomalous transport without itself being the dominant dissipative channel.

The perturbation analysis leads to the same conclusion from a complementary direction. The diffusion matrix is assembled from the same reaction currents that determine the deterministic charge transfer, so a single microscopic event generates the required correlated forcing of the inflaton, radiation, axial densities, and gauge-helicity variable. The resulting covariance evolution is numerically stable under changes of spectral estimator, initialization depth, response-sector initial covariance, and hard-helicity normalization. The gauge-induced displacement therefore does not originate from numerical differentiation, a special initialization, or an arbitrary normalization of the helicity variable. It is a property of the enlarged transport model.

There is, however, a limitation more fundamental than numerical precision. The warm condition $T/H>1$, the inflationary hierarchy, and the operator-scale hierarchy can all remain well satisfied while the transport sector lies outside parametric linear-response control. The hard-soft scale separation is modest, the gauge response is not instantaneously equilibrated, the Bose-positivity measure is violated over part of the scan, and the relevant affinities become large. At finite affinity, the net drift, susceptibility, equation of state, event statistics, and frequency-dependent noise kernel are independent real-time QCD data. Consequently, the displayed spectra are well-defined numerical predictions within the stated local linear-response closure, but they are not yet parameter-free finite-affinity $SU(3)$ predictions.

This limitation also points directly to the next theoretical target. The missing microscopic inputs can be summarized by ${\bm L}(T,\dot\phi/f,\{\mu_{5f}\},\mu_h)$, ${\bm\chi}(T,\{\mu_{5f}\},\mu_h)$, and the frequency- and momentum-dependent noise kernel ${\bm D}(\omega,\mathbf{k};T,\dot\phi/f,\{\mu_{5f}\},\mu_h)$. A real-time determination of these quantities would decide whether gauge screening merely deforms the viable SMWI trajectory, selects a narrower dynamical attractor, or qualitatively changes the existence of the strong warm branch. It would also determine the finite-affinity relation between drift and fluctuations that the present Gaussian closure necessarily assumes.

The observable interpretation is correspondingly specific. Our scalar spectrum is computed within the enlarged transport model, whereas $r_{\rm vac}$ contains only the vacuum tensor contribution because the finite-temperature transverse-traceless QCD stress and its unequal-time correlator have not been calculated. The plotted comparison with tensor constraints is correspondingly measure rather than a complete likelihood test.

The broader conclusion does not depend on any single benchmark: \emph{gauge helicity is not merely a correction to the WI friction coefficient; it can be part of the infrared dynamical system itself}. In SMWI this additional response can materially screen QCD topological dissipation and move the primordial trajectory. Precision cosmology therefore turns the problem around: obtaining a genuinely parameter-free SM prediction now requires finite-affinity, real-time non-Abelian transport and noise. The cosmological calculation identifies both the importance of that physics and the specific microscopic observables that must be computed next.
\section*{Acknowledgments}
We thank Umang Kumar, Rudnei O.~Ramos and Jaime Calderón for helpful comments and discussions on the manuscript. We are grateful to Mehrdad Mirbabayi for discussions on the effective-theory interpretation of the gauge-helicity response. A.~G. acknowledges support
from the Royal Society, UK, Fellowship funding reference: NIF\ R1\ 253963. R.H.D. acknowledges the support of the Abdus Salam International Centre for Theoretical Physics (ICTP) through its Postgraduate Diploma Programme's fellowship.
\appendix
\section{A Derivation of the quasi-steady gauge screening}
\label{sec:der_quasi_steady_gauge}
Starting from Eq.~\eqref{eq:n_dot_5f_def} in the quasi-steady limit
\begin{equation}
    n_{5f}=-\frac{2J_{\rm sph}}{3H+\Gamma_{{\rm ch},f}}\,,
\end{equation}
so
\begin{equation}
    2f\sum_f\mu_{5f}=-\sum_f\frac{4fJ_{\rm sph}}{\chi_q\left(3H+\Gamma_{{ch},f}\right)}\,.
\end{equation}
Using the definition of \(\chi_q\) in Eq.~\eqref{eq:Here_we_chi_q} and \(J_{\rm sph}\) is given by Eq.~\eqref{eq:J_sph_relation} gives Eq.~\eqref{eq:2f_sum_D_q} with \(D_q\) in Eq.~\eqref{eq:main_def_D_q}.

The gauge equation is 
\begin{equation}
    3H\chi_h\mu_h=\frac{2\zeta_h}{\alpha_s}f\,\Upsilon_{\rm sph}S+\Gamma_h \chi_h\left(\alpha_s\frac{\dot\phi}{f}-\mu_h\right)\,.
\end{equation}
Therefore 
\begin{equation}\label{eq:mu_h_long_fraction}
    \mu_h=\frac{\left(2\zeta_h/\alpha_s\right)\,f\,\Upsilon_{\rm sph}\, S+\Gamma_h\chi_h\alpha_s\dot\phi/f}{\left(3H+\Gamma_h\right)\chi_h}\,.
\end{equation}
Substituting Eqs.~\eqref{eq:2f_sum_D_q} and~\eqref{eq:mu_h_long_fraction} into the definition of \(S\), Eq.~\eqref{eq:main_def_S_enters_gauge}, produces
\begin{equation}
    S\left(1+D_q+D_h\right)=\dot\phi\left[1-\zeta_h\frac{\Gamma_h}{3H+\Gamma_h}\right]\,,
\end{equation}
where 
\begin{equation}
    D_h=\frac{\zeta_h^2\Gamma_{\rm sph}}{\alpha_s^2T\chi_h\left(3H+\Gamma_h\right)}\,.
\end{equation}
This yields Eq.~\eqref{eq:S_dot_phi_ratio}. Notice that \(D_h\) is sourced by the topological transition rate, while the numerator is controlled by the perturbative helicity relaxation. The two should not be collapsed into a single phenomenological screening parameter.
\section{Perturbation basis and numerical validation}
\label{sec:perturbations}
Throughout this appendix $\pi\equiv\dot\phi$ is a background variable and should not be confused with the numerical constant $\pi$ appearing in phase-space factors. The scalar lapse perturbation is denoted $\alpha_k$, $\kappa_k$ is the perturbation of the local expansion, and ${\bm\Psi}_k$ is the radiation momentum potential in the normalization specified below. This appendix records the conventions required to reproduce the scalar covariance calculation. The released implementation uses the fixed state ordering Eq.~\eqref{eq:X_k_u_defs} at every stage of the drift, diffusion, initialization, covariance evolution, and curvature extraction. All quantities in the following formulas are expressed in reduced-Planck units, with the appropriate powers of \(M_{\rm Pl}\) restored in reported observables.

In spatially flat gauge, define \(\pi\equiv \dot\phi\) in these units. The lapse perturbation \(\alpha_k\) and the perturbation \(\kappa_k\) of the local expansion are eliminated through
\begin{align}
    \alpha_k&=\frac{\pi\delta\phi_k-{\bm \Psi}_k}{2H}\,,\\
    \kappa_k&=-\frac{1}{2H}\left(\pi Hu_k -\pi^2\alpha_k+V_{,\phi}\delta\phi_k+\delta\rho_{R,k}\right)\,.
\end{align}
The radiation-momentum variable is normalized so that its linear equation contains \(-\delta\rho_R/(3H)\) and \(-(4\rho_R/3)\alpha/H\), with this normalization. The comoving curvature perturbation extracted from the state is
\begin{equation}
    \mathcal{R}_k=\frac{H}{\pi^2+4\rho_R/3}(\pi\delta\phi_k-{\bm \Psi}_k)\,.
\end{equation}
Consequently
\begin{equation}
    \begin{aligned}
        \mathcal{P}_\mathcal{R}(k)&=\frac{k^3}{2\pi^2}\frac{H^2}{(\pi^2+4\rho_R/3)^2}\,\text{Re}\left({\bm C}^\dagger{\bm \Sigma}_k {\bm C}\right)\,,\\
        {\bm C}&= \left(\pi,\, 0,\, 0,\, -1,\, 0,\, \cdots,\, 0\right)^T\,.
    \end{aligned}
\end{equation}
The complete drift matrix is generated from the exact linear right-hand side rather than transcribed by hand; if \(\mathcal{F}_i(N,\, {\bm X})\equiv dX_i/dN\), then \(A_{ij}=\partial\mathcal{F}_i/\partial{\bm X}_j\). Since the perturbation system is linear in \({\bm X}\) at fixed background, each column can be obtained by evaluating the exact right-hand side on the corresponding basis vector. Temperature dependence is retained through \(\partial_T\Gamma_{\rm sph},\, \partial_TL_{23},\, \partial_T\chi_q,\, \partial_T\chi_h,\, \partial_T\alpha_s,\) and \(\partial_T\Gamma_{{\rm ch},f}\), including the one-loop running of \(\alpha_s\). The microscopic rates have no additional explicit \(\phi\) dependence in the fiducial model at fixed \(f\); their field dependence enters through the evolving background temperature and velocity.

For reference, in the basis of Eq.~\eqref{eq:X_k_u_defs}, the nonzero stochastic source vectors are proportional to
\begin{align}
    {\bm g}_{\rm sph}&\propto \left(0,\frac{1}{fH},\,-\frac{\pi}{f},\, 0,\, -2,\, -2,\, -2,\, -2,\, -2,\,\frac{2\zeta_h}{\alpha_s}\right)^T\,,\\
    {\bm g}_{23}&\propto \left(0,\,\frac{\alpha_s}{fH},\,-\frac{\pi\alpha_s}{f},\, 0,\, 0,\, 0,\, 0,\, 0,\, 0,\, 1\right)^T\,,
\end{align}
with one independent unit direction for each quark chirality-flip current. Their amplitudes are fixed by Eq.~\eqref{eq:mixed_state_diffusion}; no independent radiation white-noise normalization is introduced for these reaction currents.
\section{Slow-roll dynamics in WI}
\label{sec:SR-WI}
For reference, this appendix summarizes the slow-roll hierarchy relevant for the background solutions used throughout this work. The purpose is twofold. First, it makes explicit how the familiar cold-inflation slow-roll conditions are modified by dissipation. Second, it separates the cosmological slow-roll approximation from the distinct microscopic assumptions involved in the gauge-quark transport description. These notions are sometimes conflated, but in the present work they probe parametrically different aspects of the calculation. Standard derivations of the WI attractor are reviewed in Refs.~\cite{Berera:1995ie, Berera:1996fm,Berera:2008ar, Kamali:2023lzq}, while its stability conditions were analyzed systematically in Ref.~\cite{Moss:2008yb}. Recent numerical implementations are discussed in Refs.~\cite{Rodrigues:2025neh,ORamos:2025uqs}.

The homogeneous system introduced in Sec.~\ref{sec:SMWI_quark} consists of the inflaton, the radiation bath, and the Friedmann equation. Once the dissipative force can be represented locally as \(\mathcal{F}_\phi=\Upsilon\dot\phi\) the inflaton equation in Eq.~\eqref{eq:ddot_phi_main_SMWI} can be written as
\begin{equation}
    \ddot\phi+3H\left(1+Q\right)\dot\phi+V_{,\phi}=0\,,
\end{equation}
where \(Q\) is defined by Eq.~\eqref{eq:Q_very_main_def}. The radiation equation is given by Eq.~\eqref{eq:dot_rho_very_main_def}. The slow-roll regime is characterized by
\begin{equation}
    \left|\ddot\phi\right|\ll 3H\left(1+Q\right)\left|\dot\phi\right|\,,\qquad \left|\dot\rho_R\right|\ll 4H\rho_R\,,\qquad \frac{\dot\phi^2}{2}\ll V(\phi)\,,\qquad \rho_R\ll V\,.
\end{equation}
Under these conditions the background approaches the algebraic trajectory along Eq.~\eqref{eq:quasi_steady_approx_in_SMWI}
\begin{align}
    3H\left(1+Q\right)\dot\phi&\simeq -V_{,\phi}\,,\label{eq:1+Q_V_phi}\\
    3M_{\rm Pl}^2H^2&\simeq V(\phi)\,.\label{eq:3M_Pl_V}
\end{align}
Eq.~\eqref{eq:quasi_steady_approx_in_SMWI} is the quasi-steady radiation relation. The essential difference from cold inflation is apparent in Eq.~\eqref{eq:1+Q_V_phi}: the damping of the homogeneous mode is enhanced by \(1+Q\). Consequently, a potential need not be as flat as in the \(Q\rightarrow0\) limit in order to support slowly varying motion~\cite{Berera:2008ar, Kamali:2023lzq}.

It is useful  to quantify the hierarchy directly. Introducing the usual potential slow-roll parameters
\begin{equation}\label{epsilon_V_eta_V_app}
    \epsilon_V\equiv \frac{M_{\rm Pl}^2}{2}\left(\frac{V_{,\phi}}{V}\right)^2\,,\qquad \eta_V\equiv M_{\rm Pl}^2\frac{V_{,\phi\phi}}{V}\,,
\end{equation}
Eqs.~\eqref{eq:quasi_steady_approx_in_SMWI} and~\eqref{eq:3M_Pl_V} give
\begin{equation}
    \frac{\dot\phi^2}{2V}\simeq \frac{\epsilon_V}{3\left(1+Q\right)^2}\,,\qquad \frac{\rho_R}{V}\simeq \frac{Q\,\epsilon_V}{2\left(1+Q\right)^2}\,.
\end{equation}
This radiation can be thermally important while remaining subdominant in the total energy density. This is one of the characteristic features of WI: \(T>H\) does not require \(\rho_R\simeq V\)~\cite{Berera:2008ar, Kamali:2023lzq}.

The first Hubble slow-roll parameter follows from \(\dot{H}=-\left(\dot\phi^2+4\rho_R/3\right)/\left(2M_{\rm Pl}^2\right)\). On the slow-roll trajectory
\begin{equation}
    \epsilon_H\equiv -\frac{\dot{H}}{H^2}\simeq \frac{\epsilon_V}{1+Q}\,.
\end{equation}
Inflation therefore persists while
\begin{equation}\label{eq:epsilon_V_1_Q}
    \epsilon_V<1+Q\,,
\end{equation}
In the numerical calculation we do not impose Eq.~\eqref{eq:epsilon_V_1_Q} as an exact termination rule. Instead, as stated in Sec.~\ref{sec:numerical}, the full background is evolved until the geometric condition \(\epsilon_H=1\) is reached. Eq.~\eqref{eq:epsilon_V_1_Q} should therefore be understood as the slow-roll interpretation on the numerical criterion.

If the dissipative coefficient depends on both the background field and the temperature, \(\Upsilon=\Upsilon(\phi,\,T)\), the potential parameters in Eq.~\eqref{epsilon_V_eta_V_app} do not by themselves guarantee a slowly varying attractor. One introduces~\cite{Berera:2008ar, Kamali:2023lzq, Hall:2003zp}
\begin{equation}
    \beta_\Upsilon\equiv M_{\rm Pl}^2\frac{V_{,\phi}}{V}\frac{\Upsilon_{,\phi}}{\Upsilon}\,,\qquad b\equiv T\frac{V_{,\phi T}}{V_{,\phi}}\,,\qquad c\equiv T\frac{\Upsilon_{,T}}{\Upsilon}\,.
\end{equation}
The parameters $\beta_\Upsilon$, $b$, and $c$ quantify, respectively, the field dependence of the dissipation coefficient along the potential slope, the explicit thermal deformation of the inflaton force, and the temperature dependence of dissipation. They are background stability measures and should not be confused with the microscopic response-mode criteria of Sec.~\ref{sec:enters_gauge}.

A linear stability analysis of the warm attractor gives the familiar sufficient hierarchy
\begin{equation}
    \epsilon_V,\,\left|\eta_V\right|,\, \left|\beta_\Upsilon\right|\ll 1+Q\,,\qquad \left|b\right|\ll\frac{Q}{1+Q}\,,\qquad\left|c\right|<4\,,
\end{equation}
up to mild refinements that depend on the detailed thermal sector~\cite{Moss:2008yb,Kamali:2023lzq}. The first three conditions express the suppression of background time derivatives by the enhanced friction \(3H\left(1+Q\right)\). The parameter \(b\) measures explicit thermal deformation of the force \(V_{,\phi}\), whereas \(c\) determines how rapidly the dissipative source changes as the radiation temperature responds to the background. The latter conditions are therefore stability conditions for the coupled inflaton-radiation system rather than simply statements of the flatness of \(V\).

For the potential used in this work in Eq.~\eqref{eq:Lagrangian_intro}, Eq.~\eqref{epsilon_V_eta_V_app} reduces to
\begin{equation}
    \epsilon_V=\frac{8M_{\rm Pl}^2}{\phi^2}\,, \qquad\eta_V=\frac{12M_{\rm Pl}^2}{\phi^2}=\frac{3}{2}\epsilon_V\,.
\end{equation}
Within the potential approximation adopted in Eq.~\eqref{eq:Lagrangian_intro}, \(V\) has no explicit temperature dependence and hence \(b=0\). The dissipative sector is more subtle. Already in the quark-only system, \(\Upsilon_q\), contains the screening factor \(D_q\) of Eq.~\eqref{eq:main_def_D_q}; after gauge helicity is included, the reduced background instead contains
\begin{equation}
    \Upsilon_{\rm tot}=\Upsilon_{\rm sph}\frac{S}{\dot\phi}+\Upsilon_{23}+\Upsilon_{\rm dir}\,,
\end{equation}
with \(S/\dot\phi\) determined by Eq~\eqref{eq:S_dot_phi_ratio}. Hence the effective temperature and field dependence is not described by a single monomial \(\Upsilon\propto T^c\phi^p\): it inherits the temperature dependence of \(\Gamma_{\rm sph}\), the running of \(\alpha_s\), quark chirality relaxation, and the gauge response through \(D_h\) and \(R_h\).

When useful, one may nevertheless define local logarithmic slopes along the reduced response manifold
\begin{equation}
    \beta_{\rm eff}\equiv M_{\rm Pl}^2\left.\frac{V_{,\phi}}{V}\frac{\partial\,\ln\Upsilon_{\rm tot}}{\partial\phi}\right|_{T}\,,\qquad c_{\rm eff}\left.\equiv \frac{\partial\,\ln\Upsilon_{\rm tot}}{\partial\,\ln\,T}\right|_{\phi}\,.
\end{equation}
These quantities are useful measures, but we do not use them in the full numerical scan of Sec.~\ref{sec:numerical}. The background normalization and perturbation spectrum are obtained from the dynamical system itself, in the same spirit as recent high-numerical treatments of WI~\cite{Rodrigues:2025neh,ORamos:2025uqs}. This distinction is particularly important away from a simple  power-law dissipative regime.

Combining Eq.~\eqref{eq:quasi_steady_approx_in_SMWI} with \(\rho_R=\pi^2g_*T^4/30\) gives
\begin{equation}\label{eq:temperature_slow_roll}
    T^4\simeq \frac{15}{\pi^2g_*}\frac{Q\epsilon_V}{\left(1+Q\right)^2}V\,.
\end{equation}
or equivalently
\begin{equation}
    \left(\frac{T}{H}\right)^4\simeq \frac{135}{\pi^2g_*}\frac{Q\epsilon_V}{\left(1+Q\right)^2}\frac{M_{\rm Pl}^4}{V}\,.
\end{equation}
The fourth power of the reduced Planck mass follows from dividing Eq.~\eqref{eq:temperature_slow_roll} by \(H^4\simeq V^2/(9M_{\rm Pl}^4)\); it is required for dimensional consistency. The condition \(T/H>1\) identifies the regime in which a thermal bath exists at a scale above the Hubble expansion and thermal effects can compete with the vacuum fluctuations. It is, however, not itself a slow-roll condition and, importantly for the present work, does not establish microscopic locality or linear response.

The number of \(e\)-folds between a pivot field value \(\phi_*\) and the end of inflation follows from Eq.~\eqref{eq:1+Q_V_phi}
\begin{equation}
    N_*\equiv \int_{t_*}^{t_{\rm end}}H\,dt\simeq \frac{1}{M_{\rm Pl}^2}\int_{\phi_{\rm end}}^{\phi_*}d\phi\,\frac{V}{V_{,\phi}}\left[1+Q(\phi)\right]\,.
\end{equation}
For the potential  given in Eq.~\eqref{eq:Lagrangian_intro}
\begin{equation}\label{eq:N_*_simeq_for_quartic}
    N_*\simeq\frac{1}{4M_{\rm Pl}^2}\int_{\phi_{\rm end}}^{\phi_*}d\phi \,\phi\left[1+Q(\phi)\right]\,.
\end{equation}
Because \(Q\) varies substantially along portions of the gauge+quark trajectory, Eq.~\eqref{eq:N_*_simeq_for_quartic} is not replaced by a constant-\(Q\) analytic estimate in our numerical analysis. We instead integrate the background to \(\epsilon_H=1\), yielding the \(N_*\) values. This is also why the same quartic potential can correspond to different amplitude-normalized trajectories in the quark-only and gauge-quark cases.

The slow-roll hierarchy above should be distinguished from the microscopic validity conditions studied in the main analysis. In ordinary WI the existence of a slowly varying background is often discussed together with the assumption that the bath can be integrated out into a local dissipative coefficient. In the present gauge-quark system these are logically separate statements. Only the statement of \(\epsilon_H\ll1\) is the cosmological slow-roll condition. Both \(T/H>1\) and \(T/\Lambda_{\rm EFT}\ll1\) conditions characterize the thermal and operator hierarchies discussed in Sec.~\ref{sec:numerical}. The response-relaxation hierarchy is diagnosed by the instantaneous spectrum in Fig.~\ref{fig:relaxation}, while the numerical accuracy of the homogeneous quasi-steady reduction is tested independently by the full-versus-reduced attractor comparison in Fig.~\ref{fig:full-vs-reduced}. The condition \(\left|A_r\right|\ll1\) tests the small-affinity regime associated with \(A_{\rm sph}\) and \(A_{23}\), defined in Eqs.~\eqref{eq:A_sph_event} and~\eqref{eq:A_23_main_def}, while \(N_r\gg1\) is tested by the reaction counts shown in Fig.~\ref{fig:event-counts}.

That separation is important when interpreting the numerical results. The numerical trajectories can satisfy \(\epsilon_H<1\), \(T/H>1\), and the hierarchy \(T/\Lambda_{\rm EFT}\ll1\) without implying that the finite-density \(SU(3)\) transport kernel is within its linear-response regime. Here the slow-roll relations derived above are used as background intuition and as initialization checks, whereas the central conclusion of this work is formulated in terms of microscopic transport control rather than slow roll alone.
\bibliographystyle{JHEP}
\bibliography{main}
\end{document}